\documentclass[reqno]{amsart}

\usepackage{mathtools,amsmath,amsthm,amssymb,amsfonts,enumerate,graphicx,color,pstricks,rotating}
\usepackage[T1]{fontenc} 
\numberwithin{equation}{section} 
\theoremstyle{plain}
\newtheorem{theo+}           {Theorem}      [section]
\newtheorem{prop+}  [theo+]  {Proposition}
\newtheorem{coro+}  [theo+]  {Corollary}
\newtheorem{lemm+}  [theo+]  {Lemma}
\newtheorem{defi+}  [theo+]  {Definition}
\newtheorem{conj+}  [theo+]  {Conjecture}

\theoremstyle{definition}
\newtheorem{rema+}  [theo+]  {Remark}
\newtheorem{prob+}  [theo+]  {Problem}
\newtheorem{exam+}  [theo+]  {Example}

\newenvironment{theorem}{\begin{theo+}}{\end{theo+}}
\newenvironment{proposition}{\begin{prop+}}{\end{prop+}}
\newenvironment{corollary}{\begin{coro+}}{\end{coro+}}

\newcommand{\om}{\omega}
\newcommand{\tha}{\theta}
\newcommand{\ti}{\textup i}

\newcommand{\id}{\operatorname{id}}

\newcommand{\ch}{{\operatorname{ch}}}

\begin{document}

\baselineskip 18pt
\larger[2]
\title[Special polynomials related to the eight-vertex model]
{Special polynomials related to the\\ supersymmetric eight-vertex model:\\ A summary} 
\author{Hjalmar Rosengren}
\address
{Department of Mathematical Sciences
\\ Chalmers University of Technology and University of Gothenburg\\SE-412~96 G\"oteborg, Sweden}
\email{hjalmar@chalmers.se}
\urladdr{http://www.math.chalmers.se/{\textasciitilde}hjalmar}

\begin{abstract}
We introduce and study  symmetric polynomials, which as very special cases include polynomials related to the supersymmetric eight-vertex model, and other elliptic lattice models with  $\Delta=\pm 1/2$. 
There is also a close relation to 
  affine Lie
algebra characters. After a natural change of variables, our polynomials satisfy a non-stationary Schr\"odinger equation with elliptic potential, which is related to the
 Knizhnik--Zamolodchikov--Bernard equation and to the canonical quantization of Painlev\'e VI. Moreover, specializations of our polynomials can be identified with tau functions of Painlev\'e VI, obtained from one of Picard's algebraic solutions by acting with a four-dimensional lattice
of B\"acklund transformations. In the present work, our results on these topics are summarized with a minimum of technical details.
 \end{abstract}

\maketitle

\section{Introduction}

In the present work, we introduce and study
 certain  symmetric polynomials.
Numerous special cases have appeared in connection with elliptic solvable lattice models, at
the special parameter values usually denoted  $\Delta=\pm 1/2$. 
We show that, up to a change of variables, our polynomials satisfy a Schr\"odinger equation with elliptic potential and that specializations of the polynomials are tau functions of Painlev\'e VI. For very special cases, these properties have been conjectured by Bazhanov and Mangazeev 
\cite{bm1,bm2,mb}. The rigorous proofs of our main results contain many  technical steps. For this reason, we have chosen to write the present summary, 
where the main results and ideas are described with a minimum of details.
Complete proofs
can be found in our series of preprints \cite{r1a,r1b,r2}.

The  values $ \Delta=\pm 1/2$ are truly exceptional. An intriguing feature is that exact results can be obtained not only in the infinite lattice limit but already on finite lattices.
One explanation why  $\Delta=-1/2$ is special comes from the limit to the massive sine-Gordon model, when it becomes a condition for supersymmetry \cite{fs}. Recently, Hagendorf and Fendley \cite{hf} implemented this supersymmetry on the finite lattice. Thus, we  refer to  
$ \Delta=-1/2$ as the \emph{supersymmetric} case. 

The six-vertex model with $\Delta=1/2$ contains the combinatorial ice model, where all states have equal weight. This was used by Kuperberg \cite{ku} in his simple proof of the alternating sign matrix theorem, which enumerates the states with domain wall boundary conditions.
The combinatorics of the  XXZ  and six-vertex models  at $\Delta=-1/2$ is also very rich, see \cite{zs} for a  survey. 
It seems quite interesting
  to extend results in this area to the elliptic regime, but so far only the first steps have been taken. 

In \cite{bm1}, Bazhanov and Mangazeev found that the ground state eigenvalue of Baxter's $Q$-operator for the supersymmetric periodic XYZ chain of odd length can be expressed in terms of certain polynomials, which  appear to have positive integer coefficients and thus call for a combinatorial interpretation. 
The papers \cite{mb} and \cite{ras}
 deal with ground state eigenvectors of the Hamiltonian for the same  chain. Certain components of these eigenvectors, as well as certain sums of components, again seem to be described by polynomials with positive coefficients. The same polynomials  appear for
other supersymmetric spin chains \cite{bh,fh,h}. A  rigorous investigation of the  supersymmetric XYZ  chain
was recently initiated by  Zinn-Justin \cite{zj}.

As was noted in  \cite{mb}, there are striking parallels between the
work outlined above and our previous investigation of the 8VSOS and three-colour models \cite{r0,r}. Just as the six-vertex  model contains the combinatorial ice model when $\Delta=1/2$, the corresponding combinatorial specialization of the
 8VSOS model is the three-colour model.
Extending Kuperberg's work to the elliptic regime, we expressed the domain wall partition function for the three-colour model in terms of certain special polynomials, which again conjecturally have positive integer coefficients.

In the the present work, we
 explain the relations between various polynomials introduced in \cite{bm1,mb,r,zj}, by identifying them as special cases of
a more general family of functions.  
We stress that, although  the underlying physical models are  closely related \cite{b,bv}, it is not clear why  objects as different as domain wall partition functions, eigenvalues of the $Q$-operator and eigenvectors of the Hamiltonian 
 should  lead to related special functions. 

To be more precise, we define for each non-negative integer $m$ 
a four-dimensional
lattice  $T_n^{(\mathbf k)}$ of symmetric rational functions 
in $m$ variables,
depending also on a parameter $\zeta$. 
The indices $n\in \mathbb Z$ and $\mathbf k=(k_0,k_1,k_2,k_3)\in\mathbb Z^4$ satisfy $|\mathbf k|+m=2n$. 
Since the denominator  is elementary,  $T_n^{(\mathbf k)}$ are essentially symmetric polynomials.

After a natural change of variables,
  $T_n^{(\mathbf k)}$  is a multivariable theta   function, which is closely related to
 affine Lie algebra characters.  In fact, the theta function corresponding to $T_n^{(0,0,0,0)}$ can be identified with a character of the affine Lie algebra
of type  $C_{2n}^{(1)} $. If all the indices $k_i$ are non-negative,  $T_n^{(\mathbf k)}$
is obtained from  $T_{n}^{(0,0,0,0)}$ by specializing some
of the variables to half-periods.  In the general case, it can be obtained from the character $T_{n+\sum_i\max(-k_i,0)}^{(0,0,0,0)}$
through a slightly more complicated procedure. This link to affine Lie algebras is not used to obtain our  results, but may provide some explanation for the ubiquity of the functions $T_n^{(\mathbf k)}$ in the context of solvable models.

Our first main result is that  $T_n^{(\mathbf k)}$ satisfies
a certain algebraic differential equation, see
 Theorem \ref{pdet}. Special cases  have been obtained by Bazhanov and Mangazeev \cite{bm1,mb} (without complete proof) and  Zinn-Justin \cite{zj}.  Up to a change of variables, this equation is
 a non-stationary Schr\"odinger equation
with elliptic potential, see Theorem \ref{semt}. 
When $m=1$,  it takes the form
\begin{equation}\label{nse}\psi_t=\frac 12\,\psi_{xx}-V\psi, \end{equation}
where  $V$ is the Darboux potential \cite{d,v}
\begin{equation}\label{dp}V(x,t)=\sum_{j=0}^3\frac{k_j(k_j+1)}{2}\,\wp(x-\gamma_j|1,2\pi\ti t), \end{equation}
with $\gamma_j$  the four half-periods of the $\wp$-function.
The $m$-variable case is simply the equation for $m$ non-interacting
particles with the same potential.

The equation \eqref{nse} has appeared in the literature in several contexts.
It is a canonical quantization
of  Painlev\'e VI, and has been studied from this viewpoint by Nagoya \cite{na,na2}, Suleimanov \cite{su,su2} and Zabrodin and Zotov \cite{z}, see
also \cite{cd,no,zos}. 
Under some extra condition on the parameters, it is the one-dimensional case of the Knizhnik--Zamolodchikov--Bernard heat equation satisfied by conformal blocks of Wess--Zumino--Witten theory on a torus \cite{be,ek}. 
The general case also appears in conformal field theory \cite{fl}.
More precisely, \eqref{nse} corresponds to a theory with 
central charge $c=1$,  a case  known to have close connections to Painlev\'e VI, see e.g.\ \cite{er,gil}. 
Recently, Kolb  \cite{ko} identified
the corresponding Schr\"odinger operator with the radial part of the Casimir operator for the affine Lie algebra $\widehat{sl}_2$. 
Finally, 
we mention the  paper \cite{lt}, where a more general equation, representing interacting particles, is used to study the Inozemtsev model.

An important application of the Schr\"odinger equation is that it implies bilinear relations for $T_n^{(\mathbf k)}$, see Theorem \ref{rt}. This is used to obtain our next main result, Theorem \ref{trt}, where the functions
$T_n^{(\mathbf k)}$ with $m=0$ are identified with tau functions of Painlev\'e VI, obtained from one of Picard's solutions by acting with a four-dimensional lattice of B\"acklund transformations. For particular lines in the lattice, this has been conjectured by
 Bazhanov and Mangazeev \cite{bm2}.  As an application, 
we obtain a new  quadratic differential equation for $T_n^{(\mathbf k)}$ when $m=0$, see
Proposition \ref{qdp2}.
Note that 
our tau functions can be obtained from
$m=1$ instances of $T_n^{(\mathbf k)}$, that is, from solutions
to \eqref{nse}, by specializing the variable to a half period. 
A similar observation was made in \cite{na2} for another class of solutions. Presumably, this phenomenon is linked to the 
relation between \eqref{nse}  and the Lax representation of Painlev\'e VI described in \cite{cd,su,z}.

In the final \S \ref{cns}, we explain the relation between  $T_n^{(\mathbf k)}$
and various polynomials introduced in \cite{bm1,mb,r,zj}
and also occurring in \cite{bm2,bh,fh,h,ras}.
To be precise, for the polynomials related to eigenvalues of
the $Q$-operator  and to domain wall partition functions, these relations have been rigorously proved. However, for polynomials related to eigenvectors of the Hamiltonian, the identification with our polynomials 
$T_n^{(\mathbf k)}$ is still based  on empirical observation.
A partially rigorous result exists only for the
  ``sum rule'' giving the square norm of the eigenvector, which was
recently proved by  Zinn-Justin \cite{zj}, assuming a certain conjecture.

\section{Definition and fundamental properties of $T_n^{(\mathbf k)}$}

\subsection{Notation}  
Throughout the paper,
$$\omega=e^{2\pi\ti/3}. $$
We fix
 $\tau$ in the upper half-plane, and write
 $p=e^{\pi\ti\tau}$. 
The four half-periods in  $\mathbb C/(\mathbb Z+\tau\mathbb Z)$ will be denoted
$$
\gamma_0=0,\qquad \gamma_1=\frac\tau 2,\qquad \gamma_2=\frac\tau2+\frac 12,\qquad \gamma_3=\frac 12. $$
 We will use the notation
$$\theta(x;p)=\prod_{j=0}^\infty(1-p^jx)\left(1-\frac{p^{j+1}}x\right).$$
Repeated variables are used as a short-hand for products; for instance,
$$\theta(a,b^\pm;p)=\theta(a;p)\theta(b;p)\theta(b^{-1};p). $$
Finally,  the Vandermonde product is denoted
$$\Delta(x)=\Delta(x_1,\dots,x_n)=\prod_{1\leq i<j\leq n}(x_j-x_i). $$

\subsection{Anti-symmetric theta functions} \label{sts}

For $n$ a non-negative integer, we denote
 by $\Theta_n$ the space of entire functions $f$ such that
\begin{subequations}\label{vde}
\begin{equation}\label{fqp}f(z+1)=f(z),\qquad  f(z+\tau)=e^{-6\pi \ti n(\tau+2z)}f(z),\qquad f(-z)=-f(z),\end{equation}
\begin{equation}\label{fe}f(z)+f\left(z+\frac 13\right)+f\left(z-\frac 13\right)=0. \end{equation}
\end{subequations}
This space has dimension $2n$, an explicit basis being
\begin{equation}\label{aca}f_j(z)=e^{2\pi\ti(j-3n)z}\theta(-p^{2j}e^{12\pi\ti n z};p^{12n})-e^{2\pi\ti(3n-j)z}\theta(-p^{2j}e^{-12\pi\ti n z};p^{12n}),\end{equation}
where $1\leq j\leq 3n-1$ and $3\nmid j$.

We are interested in the one-dimensional space $\Theta_n^{\wedge 2n}$,
which we realize as a space of anti-symmetric functions. 
One way to construct an element of this space is
as the alternant
\begin{equation}\label{alt}\det_{1\leq i\leq 2n,\,1\leq j\leq 3n-1,\,3\nmid j}(f_j(z_i)). 
\end{equation}
As we explain in \S \ref{afs}, \eqref{alt} is essentially a character for the
affine Lie algebra of type $C_{2n}^{(1)}$. However, for our purposes a  more useful
generator of the same space is  
\begin{multline}\label{eik}\prod_{j=1}^{2n}e^{-2\pi\ti z_j}\tha(e^{4\pi \ti z_j};p^2)
\prod_{i,j=1}^ne^{-6\pi\ti z_{n+j}}\tha(e^{6\pi \ti (z_{n+j }\pm z_i)};p^6)\\
\times\det_{1\leq i,j\leq n}\left(\frac{e^{-2\pi\ti z_{n+j}}\tha(e^{2\pi \ti (z_{n+j }\pm z_i)};p^2)}{e^{-6\pi\ti z_{n+j}}\tha(e^{6\pi \ti (z_{n+j }\pm z_i)};p^6)}\right).
 \end{multline}

The function \eqref{eik} is analogous to the 
domain wall partition function
 for the six-vertex model \cite{ick,ok,st}, 
which for $\Delta=1/2$ can be expressed as the Schur polynomial
\begin{equation}\label{sp}Z=s_{(n-1,n-1,\dots,1,1,0,0)}(z_1,\dots,z_{2n}).
\end{equation}
The usual determinant formula for this polynomial can be written
$$Z=\frac{\det_{1\leq i\leq 2n,\,1\leq j\leq 3n-1,\,3\nmid j}(z_i^{j-1})}{\Delta(z)}, $$
whereas the  Izergin--Korepin formula gives
$$Z=\frac{\prod_{i,j=1}^n(z_{n+j}^3-z_i^3)}{\Delta(z)}\det_{1\leq i,j\leq n}\left(\frac{z_{n+j}-z_i}{z_{n+j}^3-z_i^3}\right). $$
After multiplication by $\Delta(z)$, 
these expressions are clearly analogous to \eqref{alt} and \eqref{eik}. Even more to the point, 
the case $p=0$ of \eqref{eik} is essentially the Tsuchiya determinant \cite{ts} (with $\Delta=1/2$), which is the  partition function for the  six-vertex model on a rectangle bounded by one reflecting edge and three domain walls.  
Recently, Filali generalized this to the elliptic level, interpreting
 \eqref{eik} with general $p$ as a partition function for  the 8VSOS model
\cite{f}. The genuine domain wall partition function for the 8VSOS model is more complicated \cite{r0}.

\subsection{Uniformization}

With a slight modification of the notation used in \cite{r},
we will write
\begin{align}\nonumber x(z)&=\frac{\theta(- p\omega;p^2)^2\theta(\omega e^{\pm 2\pi \ti z};p^2)}{\theta(-\omega;p^2)^2\theta( p\omega e^{\pm 2\pi \ti z};p^2)},\\
\label{z}\zeta(\tau)&=\frac{\omega^2\theta(-1,- p\omega;p^2)}{\theta(- 
p,-\omega;p^2)}. \end{align}
The function $x$ generates the field of
 even elliptic function with periods $1$ and $\tau$. Moreover, 
as we discuss in \S \ref{ms}, $\zeta$ generates the field of modular functions for the  group $\Gamma_0(6,2)\simeq\Gamma_0(12)$. 
Thus, if a function of $(z,\tau)$ has the appropriate elliptic and modular behaviour, it is automatically a rational function of $(x,\zeta)$. 
We refer to the change of variables from $(z,\tau)$ to $(x,\zeta)$ as \emph{uniformization}.

Uniformizing the determinant
\eqref{eik}, we are led  to define
\begin{equation}\label{tn} T(x_1,\dots,x_{2n})
=\frac{\prod_{i,j=1}^nG(x_i,x_{n+j})}{\Delta(x_1,\dots,x_n)\Delta(x_{n+1},\dots,x_{2n})}\,\det_{1\leq i,j\leq n}\left(\frac{1}{G(x_i,x_{n+j})}\right),
 \end{equation}
where
\begin{equation}\label{g} G(x,y)=(\zeta+2)xy(x+y)-\zeta(x^2+y^2)-2(\zeta^2+3\zeta+1)xy+\zeta(2\zeta+1)(x+y).\end{equation}
Then,  $T$ is a symmetric polynomial in $2n$ variables, depending also as 
a polynomial on the parameter $\zeta$.
Moreover,
the space 
$\Theta_n^{\wedge 2n}$ is spanned by
\begin{equation}\label{tut}\prod_{j=1}^{2n}e^{-2\pi\ti z_j}\tha(e^{4\pi\ti z_j};p^2)
\tha(p\om e^{\pm 2\pi\ti z_j};p^2)^{3n-2}\Delta(x_1,\dots,x_{2n})\,T(x_1,\dots,x_{2n}), \end{equation}
where $x_j=x(z_j)$. 
The function $T$ essentially agrees with the function $H_n$  of Zinn--Justin \cite{zj}, see \eqref{thr}.

\subsection{The functions $T_n^{(\mathbf k)}$}

Let $\xi_j=x(\gamma_j)$ be the values of $x$ at the half-periods. 
Explicitly,
$$
\xi_0=2\zeta+1,\qquad \xi_1=\frac{\zeta}{\zeta+2},
\qquad \xi_2=\frac{\zeta(2\zeta+1)}{\zeta+2},
\qquad \xi_3=1.
$$
Many functions related to  solvable models can be 
 obtained by specializing
some variables of the functions $T$ to the values $\xi_j$.
As a preliminary definition, let 
\begin{equation}\label{tnkp}T_n^{(\mathbf k)}(x_1,\dots,x_m)=T(x_1,\dots,x_m,\boldsymbol\xi^{\mathbf k}), \end{equation}
 where $\mathbf k=(k_0,k_1,k_2,k_3)$ and 
$$\boldsymbol\xi^{\mathbf k}=(\underbrace{\xi_0,\dots,\xi_0}_{k_0},\underbrace{\xi_1,\dots,\xi_1}_{k_1},\underbrace{\xi_2,\dots,\xi_2}_{k_2},\underbrace{\xi_3,\dots,\xi_3}_{k_3}).$$
 Here, $m$ and $k_j$ 
are non-negative integers restricted by $m+|\mathbf k|=2n$.

We need to relax the condition that  $k_j\geq 0$. 
To motivate our extension, 
note that specializing a variable to $\gamma_j$ in the alternant
\eqref{alt} leads to a function in $\Theta_n^{\wedge(2n-1)}$ that
vanishes if one of the variables equals $\gamma_j$. Thus,
as a function of each variable, $T_n^{(\mathbf k)}$ uniformizes a function that satisfies
\eqref{vde} together with  vanishing conditions at $\gamma_j$.
For instance, if $j=0$, each element in this space has a Taylor expansion
of the form
$$f(z)=\sum_{N=k_0}^\infty b_Nz^{2N+1}. $$
Let $\Phi$ be the $1/3$-periodic function  $\Phi(z)=\theta(e^{\pm 6\pi\ti z};p^6)^{-k_0}$. Using \eqref{vde}, it is easy to see that
$$\Phi(z)\left(f\left(z+\frac 13\right)+f\left(-z+\frac 13\right)\right)=\sum_{N=-k_0}^\infty c_Nz^{2N}. $$
This leads to the idea that, to extend \eqref{tnkp} to $k_j<0$, one should impose vanishing conditions at $\xi_j$ after first applying the map $(\sigma f)(z)= f(z+1/3)+f(-z+1/3)$.

It is thus natural to look at the polynomial
\begin{multline}\label{ttv}T(x_1,\dots,x_k;x_{k+1},\dots,x_{2n})\\
=\frac{(\id^{\otimes k}\otimes\,\hat\sigma^{\otimes (2n-k)})\Delta(x_1,\dots,x_{2n})T(x_1,\dots,x_{2n})}{\Delta(x_1,\dots,x_k)\Delta(x_{k+1},\dots,x_{2n})},
 \end{multline}
where $\hat\sigma$ is a uniformization of  $\sigma$ (see \cite{r1a}
for the precise definition).
By construction, it is symmetric in its first $k$ and last $2n-k$ variables.
We can obtain different determinant formulas for \eqref{ttv} by
 letting $\hat\sigma$ act on different variables in \eqref{tn}. Explicitly, 
\begin{multline}\label{tkl}
T(x_1,\dots,x_k,x_{n+1},\dots,x_{n+l};x_{k+1},\dots,x_n,x_{n+l+1},\dots,x_{2n})\\
\begin{split}&=\frac{\prod_{{1\leq i\leq k,\,l+1\leq j\leq n}}(x_{n+j}-x_i)\prod_{{k+1\leq i\leq n,\,1\leq j\leq l}}(x_i-x_{n+j})\prod_{i,j=1}^nG(x_i,x_{n+j})}{ \Delta(x_1,\dots,x_k)\Delta(x_{k+1},\dots,x_n)\Delta(x_{n+1},\dots,x_{n+l})\Delta(x_{n+l+1},\dots,x_{2n})}\\
&\quad\times\det_{1\leq i,j\leq n}(B_{ij}),
\end{split}\end{multline}
where 
\begin{align*}B_{i,j}&=\begin{cases}
\displaystyle\frac 1{G(x_i,x_{n+j})}, & 1\leq i\leq k,\ 1\leq j\leq l,\\[4mm]
\displaystyle\frac {Q(x_i,x_{n+j})}{(x_{n+j}-x_i)G(x_i,x_{n+j})}, & 1\leq i\leq k,\ l+1\leq j\leq n,\\[4mm]
\displaystyle\frac {Q(x_{n+j},x_i)}{(x_i-x_{n+j})G(x_i,x_{n+j})}, & k+1\leq i\leq n,\ 1\leq j\leq l,\\[4mm]
\displaystyle\frac {R(x_i,x_{n+j})}{G(x_i,x_{n+j})}, & k+1\leq i\leq n,\ l+1\leq j\leq n,
\end{cases}\\
Q(x,y)&=y(y-2\zeta-1)\big((\zeta+2)y-3\zeta\big)-x\big((\zeta+2)y-\zeta\big)(2\zeta+1-3y),\\
R(x,y)
&=3(\zeta+2)^2x^2y^2+\zeta(\zeta+2)(2\zeta+1)(x^2+y^2)\\&\quad-2(\zeta^2+4\zeta+1)\big((\zeta+2)xy+\zeta(2\zeta+1)\big)(x+y)\\
&\quad+4(\zeta^4+4\zeta^3+8\zeta^2+4\zeta+1)xy
+3\zeta^2(2\zeta+1)^2.
\end{align*}

As an example,  to compute 
 $T(x_1,x_2;x_3,x_4)$  one may  use
\eqref{tkl} with $k=2$, $l=0$, which gives
\begin{multline*}T(x_1,x_2;x_3,x_4)=\big((x_4-x_1)(x_3-x_2)G(x_1,x_4)G(x_2,x_3)Q(x_1,x_3)Q(x_2,x_4)\\
-(x_3-x_1)(x_4-x_2)G(x_1,x_3)G(x_2,x_4)Q(x_1,x_4)Q(x_2,x_3)\big)/(x_2-x_1)(x_4-x_3)
\end{multline*}
or with $k=l=1$, which gives
\begin{multline*}T(x_1,x_2;x_3,x_4)=(x_4-x_1)(x_3-x_2)G(x_1,x_4)G(x_2,x_3)R(x_3,x_4)\\
-G(x_1,x_2)G(x_3,x_4)Q(x_1,x_4)Q(x_2,x_3).\end{multline*}

To define $T_n^{(\mathbf k)}$ in general, we  specialize $k_j^+=\max(k_j,0)$
of the left variables and $k_j^-=\max(-k_j,0)$ of the right variables to
$\xi_j$. More precisely, we define
\begin{equation}\label{tnk}T_n^{(\mathbf k)}(x_1,\dots,x_{m})=\frac{(-1)^{\binom{|\mathbf k^-|}2}\,T(x_1,\dots,x_m,\boldsymbol\xi^{{\mathbf k}^+};\boldsymbol\xi^{{\mathbf k}^-})}{2^{|\mathbf k^-|}\prod_{i,j=0}^3G(\xi_i,\xi_j)^{k_i^-k_j^+}\prod_{j=1}^m\prod_{i=0}^3G(x_j,\xi_i)^{k_i^-}}.\end{equation}
Here, $n\in\mathbb Z$, $\mathbf k\in\mathbb Z^4$ and $m=2n-|\mathbf k|\geq 0$ (we write $|\mathbf k|=\sum_j k_j$ also when some $k_j$ are negative).
The prefactor in \eqref{tnk} has been chosen so that 
  $$T_n^{(\mathbf k+\mathbf l)}(x_1,\dots,x_m)=T_n^{(\mathbf k)}(x_1,\dots,x_m,\xi^{\mathbf l}),\qquad \mathbf l\in\mathbb Z_{\geq 0}^4. $$

To give an example,
\begin{align*}T_0^{(-2,1,0,0)}(x)&=-\frac{T(x,\xi_1;\xi_0,\xi_0)}{4G(\xi_0,\xi_1)^2G(x,\xi_0)^2},\\
&=\frac{(2\zeta+1)^2(\zeta+2)}{\zeta^2 x^3}\left((\zeta^2+\zeta+1)x(2\zeta+1-x)+\zeta(2\zeta+1)^2\right).
\end{align*}
In general, 
 $T_n^{(\mathbf k)}$ is a symmetric rational function, depending also rationally on the parameter $\zeta$.
The denominator is elementary and can be explicitly described, so  $T_n^{(\mathbf k)}$ is essentially a symmetric polynomial.

In the case $m=0$, $T_n^{(\mathbf k)}$ depends only on $\zeta$.
This case is particularly interesting, as it includes the majority of special cases arising in
statistical mechanics. Moreover, these functions can
 be identified with tau functions of Painlev\'e~VI, see \S \ref{pas}. To make it more 
clear when we are dealing with this case, we write
$$t^{(\mathbf k)}=T_{|\mathbf k|/2}^{(\mathbf k)}. $$
Here, the only restriction on $\mathbf k\in\mathbb Z^4$ is that
$|\mathbf k|$ is even.

We can now generalize \eqref{tut}. 
Let  $\Theta_n^{\mathbf k}$ be 
the space of functions
satisfying \eqref{vde}, which are analytic except for possible poles at
 $(1/6)\mathbb Z+(\tau/2)\mathbb Z$ and such that, for $j=0,1,2,3$, 
$$\lim_{z\rightarrow \gamma_j}(z-\gamma_j)^{1-2k_j}f(z)=\lim_{z\rightarrow\gamma_j}(z-\gamma_j)^2\left(f\left(z+\frac 13\right)+f\left(-z+\frac 13\right)\right)=0. $$
Then, the function
\begin{multline}\label{ukl}\prod_{j=1}^{m}\left({e^{-2\pi\ti z_j}\theta(e^{4\pi\ti z_j};p^2)\theta(\omega p e^{\pm 2\pi \ti z_j};p^2)^{3n-2}}{\prod_{l=0}^3(x_j-\xi_l)^{k_l}}\right)\\
\times \Delta(x_1,\dots,x_{m})T_{n}^{(\mathbf k)}(x_1,\dots,x_{m}) \end{multline}
 is an element of
 $(\Theta_n^{\mathbf k})^{\wedge m}$. 

It is natural to ask whether the space  $\Theta_n^{\mathbf k}$ has dimension $m$. This is a non-trivial question, but we have found that the answer is
affirmative. It is easy to see that
$$\dim(\Theta_n^{\mathbf k})=m\qquad\Longleftrightarrow\qquad T_n^{(\mathbf k)}\not\equiv 0. $$
Thus, the question is settled by the following innocent-looking fact, which is the key for obtaining all our main results.
We will comment on the proof in \S \ref{tls}. As a consequence, 
the space $(\Theta_n^{\mathbf k})^{\wedge m}$ is one-dimensional and
spanned by \eqref{ukl}.

\begin{theorem}\label{nvt}
The functions $T_n^{(\mathbf k)}$ never vanish identically.
\end{theorem}

\subsection{Modularity}\label{ms}

Some of our results
are most naturally understood in terms of modular functions. 
Recall the notation
\begin{align*}\Gamma_0(n)&=\left\{\left(\begin{matrix}a&b\\c&d\end{matrix}\right)\in \mathrm{SL}(2,\mathbb Z);\, c\equiv 0\ \operatorname{mod}\ n \right\},\\
\Gamma_0(m,n)&=\left\{\left(\begin{matrix}a&b\\c&d\end{matrix}\right)\in \mathrm{SL}(2,\mathbb Z);\, c\equiv 0\ \operatorname{mod}\ m,\  b\equiv 0\ \operatorname{mod}\ n\right\}.
\end{align*}
These groups act on the upper half-plane $\mathbb H$ by
\begin{equation}\label{mmt}A.\tau=\frac{a\tau+b}{c\tau+d},\qquad A=\left(\begin{matrix}a&b\\c&d\end{matrix}\right). \end{equation}
A meromorphic function invariant under this action is called a modular function.

                                       The function $\zeta$ is a \emph{Hauptmodul} for the modular group $\Gamma=\Gamma_0(6,2)$, that is, it 
generates the corresponding field of modular functions. 
Equivalently, $\tau\mapsto\zeta(2\tau)$ is a Hauptmodul for the isomorphic group $\Gamma_0(12)$. 
The group $\Gamma$ has six \emph{cusps}, which correspond to
particular limits of $\tau$ in $\mathbb Q\cup\{\infty\}$.
Denoting by $\overline{\mathbb H/\Gamma}$ the union of $\mathbb H/\Gamma$ and
the  cusps,  $\zeta$ extends to a bijection from
 $\overline{\mathbb H/\Gamma}$ to  $\mathbb C\cup\{\infty\}$.
The values at the cusps are
\begin{equation}\label{zcv}\zeta=-2,\quad-1,\quad-\frac 12,\quad 0,\quad 1,\quad
\infty.\end{equation}

The normalizer of $\Gamma$ in $\mathrm{SL}(2, \mathbb Z)$
is $\Gamma_0(3)$. Thus, 
$\Gamma_0(3)/\Gamma\simeq \mathrm S_3$ 
acts naturally on $\overline{\mathbb H/\Gamma}$. The cusps split into two orbits under this action. We will refer to the first orbit, corresponding to
$\zeta\in\{-2,-1/2,1\}$ as the \emph{trigonometric} cusps and the other orbit,
$\zeta\in\{-1,0,\infty\}$, as the \emph{hyperbolic} cusps.
Our functions $T_n^{(\mathbf k)}$ behave very differently at
these two types of cusps.

The  cusps are easily understood in
the context of  the supersymmetric  XYZ  chain.
In homogeneous coordinates, the coupling constants of this chain are
$$(J_x:J_y:J_z)=\left(1:-\frac \zeta{1+\zeta}:\zeta\right). $$
 Note that
$J_xJ_y+J_xJ_z+J_yJ_z=0$, which is equivalent to  $\Delta=-1/2$. We see that the trigonometric cusps correspond to the three respective conditions
$$J_x=J_y,\qquad J_x=J_z,\qquad J_y=J_z $$
and thus to the XXZ chain, whereas the  hyperbolic cusps correspond to
$$J_x=\infty,\qquad J_y=\infty,\qquad J_z=\infty, $$ 
that is, to the XY chain.

\subsection{Symmetries}\label{sss}

The lattice of polynomials $T_n^{(\mathbf k)}$ has a symmetry under the group $\mathrm S_4$, acting by  rational transformations on the variables $x_j$ and $\zeta$ and by permuting the indices $k_j$.
This  can be understood geometrically. Namely, it is generated by an
 $\mathrm S_2\times \mathrm S_2$-action corresponding to simultaneously translating the variables $x_j$ by a half-period and an 
$\mathrm S_3$-action coming from  the modular action 
of  $\Gamma_0(3)$. We state the resulting symmetry explicitly for three
generators of $\mathrm S_4$.

 \begin{proposition}\label{sscc}
The functions  $T_n^{(\mathbf k)}=T_n^{(k_0,k_1,k_2,k_3)}(x_1,\dots,x_m;\zeta)$ satisfy
\begin{align*}
\notag\lefteqn{T_n^{(k_0,k_1,k_2,k_3)}(x_1,\dots,x_{m};\zeta)}\\
\notag&=\zeta^{2n(n-1)}\prod_{j=0}^3\xi_j^{k_j(n-1)}\prod_{j=1}^{m}x_j^{n-1}\, T_n^{(k_1,k_0,k_2,k_3)}(x_1^{-1},\dots,x_{m}^{-1};\zeta^{-1})\\
&\notag=\left(\frac{\zeta-1}{\zeta+2}\right)^{n(n-1)}\\
&\notag\quad\times T_n^{(k_2,k_1,k_0,k_3)}\left(\frac{(\zeta+2)x_1-(1+2\zeta)}{1-\zeta},\dots,\frac{(\zeta+2)x_{m}-(1+2\zeta)}{1-\zeta};-\zeta-1\right)\\
&\notag=\left(\frac{\zeta+2}{\zeta(2\zeta+1)}\right)^{n(n-1)}
\prod_{j=0}^3\xi_j^{k_j(n-1)}
\prod_{j=1}^{m}{x_j^{n-1}}\\
&\quad\times T_n^{(k_1,k_0,k_3,k_2)}\left(\frac{\zeta(2\zeta+1)}{(\zeta+2)x_1},\dots,\frac{\zeta(2\zeta+1)}{(\zeta+2)x_{m }};\zeta\right).
\end{align*}
\end{proposition}

In the case $m=0$, when we use the notation $t^{(\mathbf k)}$, 
 there are two additional symmetries not contained in 
Proposition \ref{sscc}.
The  symmetry in Proposition \ref{zscc}
is not hard to understand conceptually (see \cite{r1a}), but
 the one in Proposition \ref{esp}  is more mysterious.
It follows from the relation to Painlev\'e VI, see
\S \ref{pas}, but it would be interesting to find
 a more direct proof.

\begin{proposition}\label{zscc}
With $n=|\mathbf k|/2$, we have
\begin{multline*}t^{(k_0,k_1,k_2,k_3)}\\
=\frac{(-1)^{n+1}(\zeta+2)^{2(k_1+k_2+n+2)}\,t^{(-k_0-1,-k_1-1,-k_2-1,-k_3-1)}}{12^{n+1}\zeta^{2(k_1+k_2+2n+3)}(\zeta-1)^{2(k_2+k_3+1)}(\zeta+1)^{2(k_0+k_1+2n+3)}(2\zeta+1)^{2(k_0+k_2+1)}}. \end{multline*}
\end{proposition}

To formulate the last symmetry, let
$(Y_k)_{k\in\mathbb Z}$ be the solution to the recursion
$$Y_{k+1}Y_{k-1}=2(2k+1)Y_k^2,\qquad Y_0=Y_1=1,$$
that is,
\begin{equation}\label{yre}Y_k=\begin{cases}
\prod_{j=1}^k \frac{(2j-1)!}{(j-1)!}, & k\geq 0,\\[1mm]
\frac{(-1)^{\frac{k(k+1)}{2}}}{2^{2k+1}}\prod_{j=1}^{-k-1} \frac{(2j-1)!}{(j-1)!}, & k<0.
\end{cases} \end{equation}

\begin{proposition}\label{esp}
With $n=|\mathbf k|/2$ we have
\begin{multline*}
 t^{(k_0,k_1,k_2,k_3)}=(-1)^{(k_0+k_1+n)(k_1+k_3+n)}\frac{Y_{n-k_0}Y_{n-k_1}Y_{n-k_2}Y_{n-k_3}}{Y_{k_0}Y_{k_1}Y_{k_2}Y_{k_3}}\\
\times\left(\frac{\zeta^{k_1+k_2-n}(\zeta+1)^{k_0+k_1-n}}{(\zeta-1)^{k_0+k_1-n}(\zeta+2)^{k_1+k_2-n}(2\zeta+1)^{k_1+k_3-n}}\right)^{n-1} t^{(n-k_0,n-k_1,n-k_2,n-k_3)}.
\end{multline*}
\end{proposition}

By Proposition \ref{zscc} and Proposition \ref{esp},  
 the  $\mathrm S_4$ symmetry of Proposition \ref{sscc} is enhanced to an  $\mathrm S_4\times  \mathrm S_2\times \mathrm S_2$ symmetry when $m=0$.
This is  natural from the viewpoint of Painlev\'e theory, see again \S \ref{pas}.

\subsection{Behaviour at cusps}
\label{tls}

The proof of the fundamental Theorem \ref{nvt} is based on a careful
analysis of the limit of $T_n^{(\mathbf k)}$ as $\zeta\rightarrow -2$ or, equivalently, $p\rightarrow 0$. The behaviour at the other two trigonometric
cusps follows using Proposition \ref{sscc}.

To indicate what is going on, consider the case when all $k_j\geq 0$. 
Then, 
\begin{multline*}\lim_{\zeta\rightarrow -2}\left(\frac{\zeta+2}6\right)^{(k_1+k_2)(n-1)-\delta(k_1+k_2-1) }T_n^{(\mathbf k)}(x_1,\dots,x_m)
=\big((-1)^{k_2}2^{n}3^{-k_1}\big)^{n-1}\\
\times\chi(t_1,\dots,t_m,\underbrace{1,\dots,1}_{k_0},\underbrace{-1,\dots,-1}_{k_3})\, \chi(\underbrace{1,\dots,1}_{k_1},\underbrace{-1,\dots,-1}_{k_2}),
\end{multline*}
where $\delta(n)=[n^2/4]$ and $\chi$ is the symplectic character
\begin{equation}\label{ssc}\chi(t_1,\dots,t_n)=\chi_{\left[\frac{n-1}2\right],\left[\frac{n-2}2\right],\dots,1,1,0,0}^{\mathfrak{sp}(2n)}(t_1,\dots,t_n). \end{equation}
Thus, to prove Theorem \ref{nvt} in this case it is enough to show that
\eqref{ssc} does not vanish when all variables are specialized to $1$ or $-1$. Though
it would be nice to have a  representation-theoretic proof of this fact, our proof is quite computational. We start from the Jacobi--Desnanot identity for the matrix \eqref{tkl}. In the limit $\zeta\rightarrow -2$, it implies quadratic recurrence relations for the specialized symplectic characters,
which can be used to show that they 
 never vanish. This approach
extends to the case when some $k_j<0$, with the  characters  replaced by
more complicated functions.

We have also investigated the behaviour of $T_n^{(\mathbf k)}$ at the hyperbolic cusps, which is very different. By the symmetries, it is enough to consider the case $\zeta=0$. We have proved that
$\lim_{\zeta\rightarrow 0}T_n^{(\mathbf k)}/\zeta^L$ exists and is not identically zero,
where
$$ L=\begin{cases}(k_1+k_2)(2n-k_1-k_2-1), \\
 (k_1+k_2)(2n-k_1-k_2-1)+(n+1)(n-k_0-k_3), \\
(k_1+k_2)(2n-k_1-k_2-1)+(n+1-m)(k_1+k_2-n),
\end{cases}
$$
when, respectively,
\begin{align*}  |k_1+k_2+1|&\leq m+|k_0+k_3+1|,\\
 k_1+k_2+1&\leq -(m+|k_0+k_3+1|)
,\\
 k_1+k_2+1&\geq m+|k_0+k_3+1|.
\end{align*}
This can be used to study the  behaviour of related solutions to
Painlev\'e VI at the singular points, see Corollary \ref{scc}.

\subsection{Connection to affine Lie algebras}\label{afs}

 The character formula for affine
Lie algebras is \cite[Thm.\ 10.4]{k}
\begin{equation}\label{kc}\operatorname{ch}(\Lambda)=\frac{\sum_{w\in W}\varepsilon(w)e^{w(\Lambda+\rho)-\rho}}{\prod_{\alpha>0}(1-e^{-\alpha})^{\operatorname{mult}(\alpha)}}. \end{equation}
Here, the left-hand side is the character of a highest weight module
with dominant integral weight
$\Lambda$. The sum is over the affine Weyl group and the product is over positive roots. The sign $\varepsilon$ is the determinant of the Weyl group action,
 $\rho$  the Weyl vector and $\operatorname{mult}(\alpha)$ the root multiplicity. We refer to \cite{k} for an explanation of all
these terms.

Consider the case of the affine root system $C_n^{(1)}$.  
We denote the long roots of the underlying finite
root system $C_n$ by $\pm 2e_j$ (in the notation of \cite{k}, $e_j=v_j/\sqrt 2$).
Then, the Cartan algebra $\mathfrak h^\ast$ is the complex  vector space
with basis $\Lambda_0$, $e_1$,\dots, $e_n$, $\delta$. The root sytem
$C_n^{(1)}\subseteq \mathfrak h^\ast$ is the set of non-zero elements of
the form $\pm e_j\pm e_k+m\delta$, where $m\in\mathbb Z$. A dominant integral weight is an element  of the form
$$\Lambda=\lambda_0\Lambda_0+\lambda_1e_1+\dots+\lambda_ne_n+c\delta, $$
where  $c\in \mathbb C$ and $\lambda=(\lambda_0,\dots,\lambda_n)$ is a partition, that is, a weakly decreasing sequence of non-negative integers. 

Let us write the formal exponential function on $\mathfrak h^\ast$
as $e^{\Lambda_0}=w$, $e^{e_j}=e^{-2\pi\ti z_j}$ and $e^{\delta}=p^{-2}$. Then,
the sum and product in \eqref{kc} converge for $|p|<1$.  
The Weyl group is a semi-direct 
product of $S_n$, $\{\pm 1\}^n$ and $\mathbb Z^n$ \cite[\S 6.5]{k}.
Performing the summation over $\mathbb Z^n$ and $\{\pm 1\}^n$, one may rewrite
\eqref{kc} as
\begin{multline}\label{dc} \ch(\Lambda)=\frac{w^{\Lambda_0}p^{-c}(p^{4(\lambda_0+n+1)};p^{4(\lambda_0+n+1)})_\infty^n}{(p^2;p^2)_\infty^n\prod_{j=1}^ne^{-2\pi\ti z_j}\theta(e^{4\pi\ti z_j};p^2)\prod_{1\leq j<k\leq n}e^{-2\pi\ti z_j}\theta(e^{2\pi\ti(z_j\pm z_k)};p^2)}\\
\times\det_{1\leq i,j\leq n}\Big(e^{-2\pi\ti z_i(\lambda_j+n+1-j)}\theta(-p^{2(\lambda_0-\lambda_j+j)}e^{4\pi\ti z_i(\lambda_0+n+1)};p^{4(\lambda_0+n+1)})\\
-
e^{2\pi\ti z_i(\lambda_j+n+1-j)}\theta(-p^{2(\lambda_0-\lambda_j+j)}e^{-4\pi\ti z_i(\lambda_0+n+1)};p^{4(\lambda_0+n+1)})\Big).
\end{multline}
In particular, since $\ch(0)=1$, the case $\Lambda=0$ gives the determinant form of the $C_n^{(1)}$ Macdonald identity \cite{rs}.

The determinant in \eqref{dc} agrees with \eqref{alt} if $n$ is replaced by $2n$ and
$$\lambda=(n-1,n-1,n-1,n-2,n-2,\dots,1,1,0,0). $$
(This may be compared with \eqref{sp} and \eqref{ssc}.)
We conclude that, for some multiplier $C_n(\tau)$, 
$$\prod_{j=1}^{2n}\theta(p\omega e^{\pm 2\pi\ti z_j};p^2)^{n-1}T(x_1,\dots,x_{2n})=C_n(\tau)
p^cw^{1-n}\ch (\Lambda),  $$
where $\ch(\Lambda)$ is an affine Lie algebra character  of type $C_{2n}^{(1)}$.

There are also close relations to other affine Lie algebras.
For instance, in the context of the 
three-colour model, the fundamental object is $T_n^{(0,0,0,-1)}$ rather than $T=T_n^{(0,0,0,0)}$, see \cite{r}
 and \S \ref{tcss}. This function can  in a  similar way 
be identified with a character of type $A_{2n+1}^{(2)}$. 
We plan to treat this topic in more detail in the near future.

\section{Schr\"odinger  equation}

\subsection{Elliptic Schr\"odinger equation}\label{ess}

Any element in the one-dimensional space
 $(\Theta_n^{(\mathbf k)})^{\wedge m}$ 
 satisfies a Schr\"odinger equation with elliptic potential.

\begin{theorem}\label{semt}
Let $\Psi(z_1,\dots,z_m,\tau)$ be a meromorphic function, which for fixed $\tau$ belongs to $(\Theta_n^{\mathbf k})^{\wedge m}$, and let
 \begin{multline*}\Phi=\prod_{j=1}^m\Big(\left
(e^{-3\pi \ti z_j}\tha(e^{6\pi\ti z_j};p^6)\right)^{k_0}
\tha(p^3e^{6\pi\ti z_j};p^6)^{k_1}\\
\times\tha(-p^3e^{6\pi\ti z_j};p^6)^{k_2}
\left(e^{-3\pi \ti z_j}\tha(-e^{6\pi\ti z_j};p^6)\right)^{k_3}\Big)
.
\end{multline*}
 Then, 
$$\mathcal H\Phi^{-1}\Psi=C\Phi^{-1}\Psi, $$
where, writing $z_j=x_j/3$, $\tau=2\pi\ti t/3$,
\begin{equation}\label{om}\mathcal H= -m \frac{\partial}{\partial t}+\sum_{j=1}^m \left(\frac 12\frac{\partial^2}{\partial x_j^2}-V(x_j)\right),\end{equation}
 $C$ is  independent of the variables $z_j$ and $V$ is the  potential \eqref{dp}. 
\end{theorem}

The variables $x_j$ and $t$ are only used here to write \eqref{om} in the form it usually appears in the literature; they are not related to variables $x_j$ and $t$ used elsewhere. 
Note  that  $\Psi$ is uniquely determined up to a factor depending on $\tau$; the
factor $C$  depends on this choice of normalization.
In particular, if  $C=0$ and $m=1$ we recover \eqref{nse}.

Theorem \ref{semt} follows from the  fact that
$(\Theta_n^{\mathbf k})^{\wedge m}$ is one-dimensional, which is a consequence of Theorem \ref{nvt}. 
It is thus enough to prove that this space is preserved by $\Phi\mathcal H\Phi^{-1}$, which is a straight-forward exercise.

The following result is a uniformized version of
Theorem \ref{semt}.
The special case $m=1$, $\mathbf k=(0,n,n,-1)$,  is equivalent to  \cite[Eq.\ (27)]{bm1} (given there without a complete proof, since 
it was not known at the time that $\dim\Theta_n^{(0,n,n,-1)}=1$).
The case $m=1$, $\mathbf k=(n,n,0,-1)$ was conjectured in \cite{mb};
it is in fact equivalent to the case $\mathbf k=(0,n,n,-1)$
by Proposition \ref{sscc}. 
Moreover, the case $m=2n$,  $\mathbf k=(0,0,0,0)$ is equivalent to \cite[Eq.\ (50)]{zj}.

\begin{theorem}\label{pdet}
The function $T_n^{(\mathbf k)}$ satisfies an algebraic differential equation
\begin{multline}\label{use}\left(\sum_{j=1}^m \left(a(x_j,\zeta)\frac{\partial^2}{\partial x_j^2}+b(x_j,\zeta)\frac{\partial}{\partial x_j}+c(x_j,\zeta)\right)+ d(\zeta)\frac{\partial}{\partial\zeta}+e(\zeta)\right)\\
 \Delta(x_1,\dots,x_m) T_n^{(\mathbf k)}(x_1,\dots,x_m)=0.\end{multline}
\end{theorem}

Here,
\begin{align*}a(x,\zeta)&=(x-2\zeta-1)(x-1)\big((\zeta+2)x-\zeta\big)\big((\zeta+2)x-\zeta(2\zeta+1)\big),\\
d(\zeta)&=2m\zeta(\zeta-1)(\zeta+1)(\zeta+2)(2\zeta+1).
\end{align*}
We refer to \cite{r1b} for explicit expressions for the rational functions $b$, $c$ and $e$.

Making an appropriate change of variables in Theorem \ref{semt}, it is 
in principle straight-forward to show that 
\eqref{use} holds up to a change of the constant term $e(\zeta)$. To compute 
$e(\zeta)$
is not  trivial, see \cite[\S 3.3]{r1b}, but it is crucial for obtaining the
relation to the Painlev\'e VI equation described in \S \ref{ps}.

\subsection{An application}\label{cas}

Theorem \ref{pdet} can be used to derive many bilinear identities for the functions $t^{(\mathbf k)}$.
An alternative way to derive such identities is to use the identification with
Painlev\'e tau functions discussed in \S \ref{ps}.
We illustrate the first method by  the following result, 
which aims at characterizing $t^{(\mathbf k)}$ by a very short list of properties.
This will in fact  be used in \S \ref{ps} to obtain the above-mentioned
identification with tau functions.

\begin{theorem}\label{rt}
The functions $t^{(\mathbf k)}$ satisfy the two identities
\begin{subequations}\label{km}
\begin{multline}\label{kma}
t^{(\mathbf k-2\mathbf e_0)}t^{(\mathbf k+\mathbf e_0+\mathbf e_1)}=\zeta^2(\zeta+1)(\zeta-1)(2\zeta+1)^2\\
\times\left(\frac{1}{2k_0-1}\,t^{(\mathbf k)}\frac{d t^{(\mathbf k-\mathbf e_0+\mathbf e_1)}}{d \zeta}-\frac{1}{2k_0+1}\frac{d t^{(\mathbf k)}}{d \zeta} t^{(\mathbf k-\mathbf e_0+\mathbf e_1)}\right)\\
+\frac{\zeta(2\zeta+1)}{2(2k_0-1)(2k_0+1)(\zeta+2)}A^{(\mathbf k)}t^{(\mathbf k)}t^{(\mathbf k-\mathbf e_0+\mathbf e_1)},
\end{multline}
\begin{multline}\label{kmb}
t^{(\mathbf k-2\mathbf e_0)}t^{(\mathbf k+\mathbf e_0-\mathbf e_1)}=\frac{(\zeta+1)(\zeta-1)(2\zeta+1)^2(\zeta+2)^2}{\zeta^2}\\
\times\left(\frac{1}{2k_0-1}\,t^{(\mathbf k)}\frac{d t^{(\mathbf k-\mathbf e_0-\mathbf e_1)}}{d \zeta}-\frac{1}{2k_0+1}\frac{d t^{(\mathbf k)}}{d \zeta} t^{(\mathbf k-\mathbf e_0-\mathbf e_1)}\right)\\
+\frac{(2\zeta+1)(\zeta+2)}{2(2k_0-1)(2k_0+1)\zeta^3}B^{(\mathbf k)}t^{(\mathbf k)}t^{(\mathbf k-\mathbf e_0-\mathbf e_1)},
\end{multline}
\end{subequations}
where $\mathbf e_j$ are unit vectors and
{\allowdisplaybreaks
\begin{align*}
A^{(\mathbf k)}&=(2\zeta^4-23\zeta^3-36\zeta^2-5\zeta+8)k_0^2-\zeta(2\zeta+1)(3\zeta^2+10\zeta+5)k_1(2k_0+k_1)\\
&\quad-\zeta(6\zeta^3+19\zeta^2+4\zeta-11)k_2^2-\zeta(2\zeta+1)(3\zeta^2+2\zeta+1)k_3^2\\
&\quad-2\zeta(\zeta-1)(2\zeta+1)(\zeta+3)(k_0+k_1)k_2\\
&\quad-2(\zeta-1)(2\zeta+1)(3\zeta^2+9\zeta+4)(k_0+k_1)k_3\\
&\quad-2(2\zeta+1)(\zeta^3+6\zeta^2+3\zeta-4)k_2k_3-4(2\zeta+1)(\zeta^2+5\zeta+3)(k_0+k_1)\\
&\quad+4(2\zeta+1)(2\zeta^3+5\zeta^2-\zeta-3)k_2+4(2\zeta+1)(\zeta^2+\zeta+1)k_3\\
&\quad-4(\zeta+1)^2(2\zeta^2-\zeta+2),\\
B^{(\mathbf k)}&=(10\zeta^4+13\zeta^3-28\zeta^2-41\zeta-8)k_0^2
-\zeta(2\zeta+1)(3\zeta^2+10\zeta+5)k_1(k_1-2k_0)\\
&\quad-\zeta(6\zeta^3+19\zeta^2+4\zeta-11)k_2^2-\zeta(2\zeta+1)(3\zeta^2+2\zeta+1)k_3^2\\
&\quad+2\zeta(\zeta-1)(2\zeta+1)(\zeta+3)(k_0-k_1)k_2-2(2\zeta+1)(\zeta^3+6\zeta^2+3\zeta-4)k_2k_3\\
&\quad+2(\zeta-1)(2\zeta+1)(3\zeta^2+9\zeta+4)(k_0-k_1)k_3\\
&\quad+2(\zeta-1)(2\zeta+1)(\zeta+3)(3\zeta+2)(k_1-k_0)\\
&\quad+2(2\zeta+1)(5\zeta^3+12\zeta^2-5\zeta-6)k_2+2(2\zeta+1)(3\zeta^3+8\zeta^2-3\zeta-2)k_3\\
&\quad-2(8\zeta^4+18\zeta^3-7\zeta^2-18\zeta-4).
\end{align*}
}
Moreover, the lattice of functions $t^{(\mathbf k)}$, where $\mathbf k\in\mathbb Z^4$ with $\sum_j k_j$ even, is uniquely determined by  \eqref{km}, the three  values
$$t^{(0,0,0,0)}=t^{(1,-1,0,0)}=1,\qquad
t^{(0,-1,-1,0)}=-\frac{2\zeta^2(\zeta-1)(\zeta+1)^2(2\zeta+1)}{(\zeta+2)^2} $$
and the symmetries of \emph{Proposition \ref{sscc}} (with $m=0$) and
\emph{Proposition \ref{zscc}}.
\end{theorem}

To explain the main idea, we sketch the proof of
\eqref{kma}. Applying the Jacobi--Desnanot identity
to the matrix \eqref{tkl} yields the recursion
\begin{multline*}(a-b)(c-d)T(\mathbf x;\mathbf y)T(a,b,c,d,\mathbf x;\mathbf y)
=G(a,d)G(b,c)T(a,c,\mathbf x;\mathbf y)T(b,d,\mathbf x;\mathbf y)\\
-G(a,c)G(b,d)T(a,d,\mathbf x;\mathbf y)T(b,c,\mathbf x;\mathbf y).
\end{multline*}
An appropriate specialization of the variables gives
\begin{multline*}(x-\xi_0)(\xi_0-\xi_1)t^{(\mathbf k)}T_{n+2}^{(\mathbf k+2\mathbf e_0+\mathbf e_1)}(x)
\\
=G(x,\xi_1)G(\xi_0,\xi_0)T_{n+1}^{(\mathbf k+\mathbf e_0)}(x)t^{(\mathbf k+\mathbf e_0+\mathbf e_1)}-
G(x,\xi_0)G(\xi_0,\xi_1)
t^{(\mathbf k+2\mathbf e_0)}T_{n+1}^{(\mathbf k+\mathbf e_1)}(x),
\end{multline*}
where $|\mathbf k|=2n$.
Differentiating  with respect to $x$ and  letting  $x=\xi_0$ gives
\begin{multline}\label{tpr}(\xi_0-\xi_1)t^{(\mathbf k)}t^{(\mathbf k+3\mathbf e_0+\mathbf e_1)}
\\
=\left(G(\xi_0,\xi_0)\frac{\partial G}{\partial x}(\xi_0,\xi_1)-
\frac{\partial G}{\partial x}(\xi_0,\xi_0)G(\xi_0,\xi_1)\right)t^{(\mathbf k+2\mathbf e_0)}t^{(\mathbf k+\mathbf e_0+\mathbf e_1)}\\
+G(\xi_0,\xi_0)G(\xi_0,\xi_1)\left(\frac{\partial T_{n+1}^{(\mathbf k+\mathbf e_0)}}{\partial x}(\xi_0)t^{(\mathbf k+\mathbf e_0+\mathbf e_1)}-t^{(\mathbf k+2\mathbf e_0)}\frac{\partial T_{n+1}^{(\mathbf k+\mathbf e_1)}}{\partial x}(\xi_0)\right).
\end{multline}
The  point is now that the specialized derivatives of $T$-functions can be expressed  in terms of $t$-functions using the Schr\"odinger equation.
 Indeed, since $a(\xi_0,\zeta)=0$, if we let $m=1$ and $x_1=\xi_0$
in  \eqref{use}, we get a linear relation of the form
$$A\frac{\partial T_n^{(\mathbf k)}}{\partial x}(\xi_0)+ BT_n^{(\mathbf k)}(\xi_0)+C\frac{\partial T_n^{(\mathbf k)}}{\partial \zeta}(\xi_0)=0. $$
 On the other hand, differentiating the equality $T_n^{(\mathbf k)}(\xi_0)=t^{(\mathbf k+\mathbf e_0)}$ gives
$$2\frac{\partial T_n^{(\mathbf k)}}{\partial x}(\xi_0)+\frac{\partial T_n^{(\mathbf k)}}{\partial \zeta}(\xi_0)=\frac{d t^{(\mathbf k+\mathbf e_0)}}{d \zeta}. $$
Eliminating $\partial T_n^{(\mathbf k)}/\partial\zeta$ from these two equations
gives
\begin{equation}\label{bst}\frac{\partial T_n^{(\mathbf k)}}{\partial x}(\xi_0)=Dt^{(\mathbf k+\mathbf e_0)}+E\frac{d t^{(\mathbf k+e_0)}}{d \zeta}, \end{equation}
where $D$ and $E$ are  explicit coefficients.
Using \eqref{bst} on the right-hand side of \eqref{tpr}
 gives, after replacing $k_0$ by $k_0-2$, \eqref{kma}.

\section{Painlev\'e  VI}
\label{ps}

\subsection{B\"acklund transformations}
 Painlev\'e VI is the  differential equation
\begin{align}\notag\frac{d^2q}{dt^2}&=\frac 12\left(\frac 1q+\frac 1{q-1}+\frac 1{q-t}\right)
\left(\frac{dq}{dt}\right)^2-\left(\frac 1t+\frac 1{t-1}+\frac 1{q-t}\right)\frac{dq}{dt}\\ 
\label{py}&\quad +\frac{q(q-1)(q-t)}{t^2(t-1)^2}\left(\alpha+\beta\frac t{q^2}+\gamma\frac{t-1}{(q-1)^2}+\delta\frac{t(t-1)}{(q-t)^2}\right).\end{align}
We will briefly review the rich symmetry theory of
 this equation. It is mainly due to Okamoto \cite{o}, but we  follow the  exposition of Noumi and Yamada \cite{ny}.
We introduce parameters $\alpha_0,\dots,\alpha_4$ satisfying the constraint
\begin{equation}\label{ac}\alpha_0+\alpha_1+2\alpha_2+\alpha_3+\alpha_4=1
\end{equation}
and related to the parameters of \eqref{py} by
$$\alpha=\frac{\alpha_1^2}{2},\qquad \beta=-\frac{\alpha_4^2}{2},\qquad
\gamma=\frac{\alpha_3^2}{2},\qquad \delta=\frac{1-\alpha_0^2}{2}.
 $$
We let 
\begin{multline*}H=q(q-1)(q-t)p^2-\big\{(\alpha_0-1)q(q-1)+\alpha_3q(q-t)+\alpha_4(q-1)(q-t)\big\}p\\
+\alpha_2(\alpha_1+\alpha_2)(q-t).\end{multline*}
Then, \eqref{py} is equivalent to the Hamiltonian system
\begin{equation}\label{ph}t(t-1)\frac{dq}{dt}=\frac{\partial H}{\partial p},\qquad t(t-1)\frac{dp}{dt}=-\frac{\partial H}{\partial q}. \end{equation}

The system \eqref{ph} admits many symmetries, or \emph{B\"acklund transformations}. Indeed, it is invariant under the involutions
 $s_j$,  $r_j$ and  $t_j$ 
defined as follows.
\vspace{1ex}
\begin{center}
\begin{tabular}{|c||ccccc|c c c|}
\hline
 & $\alpha_0$ & $\alpha_1$ & $\alpha_2$ & $\alpha_3$ & $\alpha_4$ & $q$ & $p$ & $t$\\
\hline \hline
&&&&&&&\\[-4mm]
$s_0$ & $-\alpha_0$ & $\alpha_1$ & $\alpha_2+\alpha_0$ & $\alpha_3$ & $\alpha_4$ & $q$ & $p-\frac{\alpha_0}{q-t}$ & $t$\\
$s_1$ & $\alpha_0$ & $-\alpha_1$ & $\alpha_2+\alpha_1$ & $\alpha_3$ & $\alpha_4$ & $q$ & $p$ & $t$\\
$s_2$ & $\alpha_0+\alpha_2$ & $\alpha_1+\alpha_2$ & $-\alpha_2$ & $\alpha_3+\alpha_2$ & $\alpha_4+\alpha_2$ & $q+\frac{\alpha_2}p$ & $p$ & $t$\\
$s_3$ & $\alpha_0$ & $\alpha_1$ & $\alpha_2+\alpha_3$ & $-\alpha_3$ & $\alpha_4$ & $q$ & $p-\frac{\alpha_3}{q-1}$ & $t$\\[1mm]
$s_4$ & $\alpha_0$ & $\alpha_1$ & $\alpha_2+\alpha_4$ & $\alpha_3$ & $-\alpha_4$ & $q$ & $p-\frac{\alpha_4}{q}$ & $t$\\[1mm]
$r_1$ & $\alpha_1$ & $\alpha_0$ & $\alpha_2$ & $\alpha_4$ & $\alpha_3$ & $\frac{t(q-1)}{q-t}$ & $\!\!\frac{(t-q)((q-t)p+\alpha_2)}{t(t-1)}$ & $t$\\[1mm]
$r_3$ & $\alpha_3$ & $\alpha_4$ & $\alpha_2$ & $\alpha_0$ & $\alpha_1$ & $\frac{t}{q}$ & $-\frac{q(pq+\alpha_2)}{t}$ & $t$\\[1mm]
$t_1$ & $\alpha_0$ & $\alpha_4$ & $\alpha_2$ & $\alpha_3$ & $\alpha_1$ & $\frac 1q$ & $-q(pq+\alpha_2)$ & $\frac 1t$\\
$t_3$ & $\alpha_0$ & $\alpha_1$ & $\alpha_2$ & $\alpha_4$ & $\alpha_3$ & $1-q$ & $-p$ & $\!\!\!\!1-t$\\
\hline
\end{tabular}
\end{center}
\vspace{1ex}

We will write $r_4=r_1r_3=r_3r_1$.
We consider the B\"acklund transformations as 
 automorphisms of the differential
 field $\mathcal F_0$ generated by $\alpha_j$, $q$, $p$ and $t$, subject to the relation \eqref{ac}, and equipped with the derivation
$$\delta=\frac{\partial H}{\partial p}\frac{\partial}{\partial q}-\frac{\partial H}{\partial q}\frac{\partial}{\partial p}+t(t-1)\frac{\partial}{\partial t}. $$
The  transformations
\begin{align*}T_1&=r_1s_1s_2s_3s_4s_2s_1,\qquad T_2=s_0s_2s_1s_3s_4s_2s_1s_3s_4s_2,\\
T_3&=r_3s_3s_2s_1s_4s_2s_3,\qquad T_4=r_4s_4s_2s_1s_3s_2s_4\end{align*}
 generate an action of $\mathbb Z^4$ on $\mathcal F_0$.

\subsection{Tau functions}
\label{tss}

 A rational solution of Painlev\'e VI can be identified with a
field automorphism $\mathbf X:\mathcal F_0\rightarrow\mathbb C(t)$ such that
$\mathbf X(t)=t$ and 
\begin{equation}\label{xid}\mathbf X\circ\delta=\delta\circ \mathbf X,\end{equation}
 where $\delta$ acts on $\mathbb C(t)$ as $t(t-1)\cdot d/dt$. Since we are interested in more general algebraic solutions, we will  define extensions $\mathcal F$ and $\mathcal M$ of the differential fields $\mathcal F_0$ and $\mathbb C(t)$, respectively, 
and consider solutions
as automorphisms $\mathbf X:\mathcal F\rightarrow \mathcal M$.
We need  $\mathcal F$ to contain \emph{tau functions}, which represent inverse logarithmic derivatives of appropriate modifications of the Hamiltonian. 
One way to do this was proposed by Masuda \cite{m}. It is, however, not ideal for our purposes and we will therefore work with a variation of Masuda's construction.

We introduce the modified Hamiltonian
\begin{align}\notag h_0&=H+\frac t{12}\left(
2(\alpha_0-1)^2-\alpha_1^2+2\alpha_3^2-\alpha_4^2+6(\alpha_0-1)\alpha_3\right)\\
\label{mh}&\quad+\frac {t-1}{12}\left(2(\alpha_0-1)^2-\alpha_1^2-\alpha_3^2+2\alpha_4^2+6(\alpha_0-1)\alpha_4\right).
\end{align}
Note that \eqref{ph} holds with $H$ replaced by $h_0$. 
We also define
$$h_1=r_1(h_0),\qquad h_3=r_3(h_0),\qquad h_4=r_4(h_0),\qquad h_2=h_1+s_1(h_1)-\frac t 3+\frac 16. $$

We  define $\mathcal F$ to be the field extension of 
  $\mathcal F_0$ by the additional generators $u$, $v$, $\tau_0,\dots,\tau_4$. The generators $u$ and $v$ satisfy
$$t=u^2v^4,\qquad 1-t=u^4v^2$$
and thus formally correspond to the
roots $t^{-1/6}(1-t)^{1/3}$ and $t^{1/3}(1-t)^{-1/6}$.
We extend $\delta$ to the new generators by
$$\delta(u)=\frac{u(t+1)}{6},\qquad \delta(v)=\frac{v(t-2)}{6},\qquad \delta(\tau_j)=\tau_jh_j,\qquad j=0,\dots,4. $$
Finally, we extend  the B\"acklund transformations to $\mathcal F$ by the 
following table.

\vspace{1ex}
\begin{center}
\begin{tabular}{|c||cc|c c c c c|}
\hline
 & $u$ & $v$ & $\tau_0$ & $\tau_1$ & $\tau_2$ & $\tau_3$ & $\tau_4$ \\
\hline \hline
&&&&&&&\\[-4mm]
$s_0$ & $u$ & $v$ & $\frac{\ti(t-q)\tau_2}{u^2v^2\tau_0} $ & $\tau_1$ & $\tau_2$ & $\tau_3$ & $\tau_4$\\
$s_1$ & $u$ & $v$ & $\tau_0$ & $\frac{\ti uv\tau_2}{\tau_1}$ & $\tau_2$ & $\tau_3$ & $\tau_4$\\
$s_2$ & $u$ & $v$ & $\tau_0$ & $\tau_1$ & $\frac{p\tau_0\tau_1\tau_3\tau_4}{\tau_2}$ & $\tau_3$ & $\tau_4$\\
$s_3$ & $u$ & $v$ & $\tau_0$ & $\tau_1$ & ${\tau_2}$ & $\frac{(1-q)\tau_2}{u\tau_3}$ & $\tau_4$\\
$s_4$ & $u$ & $v$ & $\tau_0$ & $\tau_1$ & ${\tau_2}$ & $\tau_3$ & $\frac{q\tau_2}{v\tau_4}$\\
$r_1$ & $u$ & $-v$ & $\tau_1$ & $\tau_0$ & $\frac{(q-t)\tau_2}{u^3v^3}$ & $\tau_4$ & $\tau_3$\\[1mm]
$r_3$ & $-u$ & $v$ & $\tau_3$ & $\tau_4$ & $\frac{\ti q\tau_2}{uv^2}$ & $\tau_0$ & $\tau_1$\\
$t_1$ & $u$ & $\frac{\ti}{uv}$ & $\ti\tau_0$ & $\tau_4$ & $-q\tau_2$ & $\tau_3$ & $\tau_1$\\
$t_3$ & $v$ & $u$ & $\ti\tau_0$ & $\tau_1$ & $\tau_2$ & $\tau_4$ & $\tau_3$\\
\hline
\end{tabular}
\end{center}
\vspace{1ex}

With this definition,
 the operators $T_j$ define an action of
$\mathbb Z^4$ on $\mathcal F$. This is in contrast to the alternative definition of Masuda. 
However, not all algebraic relations for B\"acklund transformations extend from $\mathcal F_0$ to $\mathcal F$. 
For instance, $t_1$ and $t_3$ generate an action of the dicyclic group of order $12$ on $\mathcal F$, but an action of $\mathrm S_3$ on $\mathcal F_0$.

An important application of tau functions is encoded in the identity
\begin{equation}\label{tqf} T_1^{l_1}T_2^{l_2}T_3^{l_3}T_4^{l_4}(q)=(-1)^{l_3+l_4}\ti uv^2\frac{\tau_{l_1,l_2,l_3,l_4+1}\tau_{l_1,l_2+1,l_3,l_4-1}}{\tau_{l_1+1,l_2,l_3,l_4}\tau_{l_1-1,l_2+1,l_3,l_4}},\end{equation}
where
\begin{equation}\label{tf}\tau_{l_1l_2l_3l_4}=T_1^{l_1}T_2^{l_2}T_3^{l_3}T_4^{l_4}\tau_0.\end{equation}
Thus, if we act on a solution of Painlev\'e VI by  $\mathbb Z^4$, the resulting new solutions will typically factor into four non-trivial parts, see
\eqref{qte} for an example.

Painlev\'e VI can be reformulated as a differential equation for the Hamiltonian, known as the $\operatorname{E_{VI}}$ equation \cite{jm,o}. 
In terms of the parameters 
$$b_1=\frac{\alpha_3+\alpha_4}{2},\quad b_2=\frac{\alpha_4-\alpha_3}2,\quad
b_3=\frac{\alpha_0+\alpha_1-1}{2},\quad b_4=\frac{\alpha_0-\alpha_1-1}{2} $$
it takes the form
\begin{multline}\label{e6}\frac{dh}{dt}\left(t(t-1)\frac{d^2h}{dt^2}\right)^2+\left(\frac{dh}{dt}\left(2h-(2t-1)\frac{dh}{dt}\right)+b_1b_2b_3b_4\right)^2\\
=\prod_{k=1}^4\left(\frac{dh}{dt}+b_k^2\right), \end{multline}
where $h$ is related to \eqref{mh} by 
$$h=h_0-\frac{C}{24}\,(2t-1), $$
with
\begin{equation}\label{c}C=(\alpha_0-1)^2+\alpha_1^2+\alpha_3^2+\alpha_4^2=2(b_1^2+b_2^2+b_3^2+b_4^2). \end{equation}

Expressing $h$ in terms of tau functions, \eqref{e6} takes a rather complicated form. To obtain a simpler identity, we first cancel the term $\prod_k b_k^2$ 
and the factor $dh/dt$
on both sides, then  differentiate in $t$ and finally cancel the factor $d^2h/dt^2$. Rewriting the result in terms of   $\tau=\tau_0$ gives
\begin{multline}\label{qd}
\delta^4(\tau)\tau-4\delta^3(\tau)\delta(\tau)+2(1-2t)\delta^3(\tau)\tau+3\delta^2(\tau)^2-2(1-2t)\delta^2(\tau)\delta(\tau)\\
-\frac{(C-6)t(t-1)+C-3}3\,\delta^2(\tau)\tau+\frac{C(t^2-t+1)-3}{3}\,\delta(\tau)^2\\
+\frac{Ct(t-1)(2t-1)}{6}\,\delta(\tau)\tau-\frac{t(t-1)G}8\,\tau^2=0,
\end{multline}
where
\begin{multline}\label{g2}
G=(\alpha_4-\alpha_3)(\alpha_3+\alpha_4)(\alpha_0+\alpha_1-1)(\alpha_0-\alpha_1-
1)t\\
+(\alpha_3-\alpha_1)(\alpha_3+\alpha_1)(\alpha_0+\alpha_4-1)(\alpha_0-\alpha_4-1
).
\end{multline}
Acting on this equation by $\mathbb Z^4$ we find that $\tau=\tau_{l_1l_2l_3l_4}$ satisfies \eqref{qd}, 
with $C$ and $G$  obtained from \eqref{c} and \eqref{g2} by replacing
each  $\alpha_j$ with $\alpha_j-l_j$. Here, $l_0$ is defined by
$$l_0+l_1+2l_2+l_3+l_4=0.$$
Somewhat surprisingly, we have not found this result in the literature.
Analogous results for other Painlev\'e equations are discussed in \cite{c}.

\subsection{An algebraic Picard solution}
When 
$$\alpha_0=\alpha_1=\alpha_3=\alpha_4=0,\qquad \alpha_{2}=\frac 12,$$
 Painlev\'e VI
can be solved explicitly in terms of Weierstrass's $\wp$-function. 
This was done by Picard already in 1889 \cite{pi}. 
The general solution is labelled by two complex parameters $\nu_1$, $\nu_2$; it is algebraic if $\nu_1,\, \nu_2\in\mathbb Q$ \cite{maz}. In \cite{bm2}, the solution with $(\nu_1,\nu_2)=(1,1/3)$ was expressed as
\begin{equation}\label{pa}q^4-4tq^3+6tq^2-4tq+t^2=0. \end{equation}
Considering this as a seed solution, we are 
 interested in the corresponding lattice of tau functions \eqref{tf}.

We make the change of variables $t\mapsto \tau$, where
$$t=\frac{\tha(-1;p^6)^4}{\tha(-p^3;p^6)^4},\qquad p=e^{\pi\ti\tau}. $$
In terms of the function \eqref{z},  
\begin{equation}\label{tz}t=\frac{\zeta(\zeta+2)^3}{(2\zeta+1)^3}.\end{equation}
Note that the cusps \eqref{zcv} 
 correspond precisely to the singular points $t=0,1,\infty$ of \eqref{py},
with one trigonometric and one hyperbolic cusp at each singular point.
Substituting \eqref{tz} into \eqref{pa},
 one finds the modular solution 
$$q=\frac{\zeta(\zeta+2)}{2\zeta+1}.$$

\subsection{Modular tau functions} 
\label{mss}

Although our
 solution is modular for $\Gamma_0(6,2)$,
 the corresponding tau functions are only modular for a  subgroup. The minimal field $\mathcal M$ containing all
 modular functions that we  need is generated by
$$\phi_1=\frac{\eta(\tau/2)^2}{\eta(\tau)^2},\quad \phi_2=\frac{\eta(2\tau)^2}{\eta(\tau)^2},\quad 
\phi_3=\frac{\eta(3 \tau/2)}{\eta(\tau/2)},\quad \phi_4=\frac{\eta(6\tau)}{\eta(2\tau)},\quad \phi_{5}=\frac{\eta(3\tau)}{\eta(\tau)}, $$
where
$$\eta(\tau)=e^{\frac{\pi\ti\tau}{12}}\prod_{k=1}^\infty(1-p^{2k\pi\ti\tau}) $$
is Dedekind's eta function. 
These functions are  modular for the group $K$, consisting of 
transformations \eqref{mmt} with
$$a\equiv d\equiv\pm 1\ \operatorname{mod}\ 12,\qquad 
b\equiv 0\ \operatorname{mod}\ 24,\qquad c\equiv 0\ \operatorname{mod}\ 72.$$
One can show that $\mathcal M$ contains  $\mathbb C(\zeta)$
and that $\phi_j^{12}\in\mathbb C(\zeta)$ for each $j$. 
Thus, with
$$\delta=t(t-1)\frac{d}{dt}=\frac{\zeta(\zeta+1)(\zeta-1
)(\zeta+2)}{2(2\zeta+1)^2}\frac{d}{d\zeta} $$
we have
$\delta(\phi_j)/\phi_j=\delta(\phi_j^{12})/12\phi_j^{12}\in\mathbb C(\zeta)$.
In particular,  $\mathcal M$ is closed under $\delta$.

The normalizer of $K$ in $\mathrm{SL}(2,\mathbb Z)$ is 
$\Gamma_0(3)$, which
 is generated by  $T(\tau)=\tau+1$  and $U(\tau)=(\tau-1)/(3 \tau-2)$.
Let
$t_1=(UT^3)^3$ and $t_3=(T^3U)^3$.
We use the same notation as for  B\"acklund transformations
in view of the following result.

\begin{proposition}\label{tep}
The following equations define  a field automorphism $\mathcal F\rightarrow\mathcal M$:
$${\bf X}(\alpha_0)={\bf X}(\alpha_1)={\bf X}(\alpha_3)={\bf X}(\alpha_4)=0, \qquad {\bf X}(\alpha_2)=\frac 12,  $$
\begin{align*}
{\bf X}(p)&=\frac{2\zeta+1}{2(1-\zeta)(\zeta+2)},& 
{\bf X}(q)&=\frac{\zeta(\zeta+2)}{2\zeta+1},& 
 {\bf X}(t)&=\frac{\zeta(\zeta+2)^3}{(1+2\zeta)^3},\\
{\bf X}(u)&=\frac{\phi_1^2\phi_3^4}{2^{2/3}\phi_5^4},& {\bf X}(v)&=-\frac{2^{4/3}\phi_2^2\phi_4^4}{\phi_5^4},\\
{\bf X}(\tau_0)&=\frac 1{\phi_5},&
{\bf X}(\tau_1)&=-\frac{\phi_3\phi_4}{\phi_5^2},&
{\bf X}(\tau_2)&=\frac{ 2^{-2/3}\ti\phi_5^4}{\phi_1^2\phi_2^2\phi_3^2\phi_4^2},\\
 {\bf X}(\tau_3)&=\frac{e^{\frac{\pi\ti}4}\phi_5}{\phi_3},
& {\bf X}(\tau_4)&=\frac{e^{\frac{3\pi\ti}4}\phi_5}{\phi_4}.
\end{align*}
This extension satisfies \eqref{xid}, as well as the identities
\begin{subequations}\label{xtsx}
\begin{align}\label{xttx}
\mathbf X\circ t_j=t_j\circ\mathbf X,\qquad j=1,\,3,\\
\mathbf X\circ s_j=\mathbf X, \qquad j=0,\,1,\,3,\,4. 
\end{align}
\end{subequations}
\end{proposition}

This gives a convenient realization of all tau functions corresponding to  \eqref{pa}
as modular functions, with the B\"acklund transformations $t_j$ acting by modular transformations.

\subsection{Identification of tau functions}
\label{its}
We are now ready to formulate the main result of \S \ref{ps},
which identifies the functions $t^{(\mathbf k)}$ with Painlev\'e
    tau functions. We need the normalizing factor
\begin{align*}\notag\phi_{l_1l_2l_3l_4}&=\frac{(-1)^{\binom{l_1+1}3+\binom{l_3+1}3+\binom{l_4+1}3+\left(\binom{l_3+1}2+l_1l_3+l_2\right)l_4}\ti^{\binom{l_3+1}2+\binom{l_4+1}2-l_1^2l_3+l_1l_4^2+l_2+l_3+l_4}}{2^{l_0(l_0-1)+l_1^2+l_3^2+l_4^2}}\\
\notag&\quad\times\zeta^{l_4^2-l_0(l_0-1)-(l_0+l_2)(l_2+l_4)}(\zeta+1)^{l_3^2-l_0(l_0-1)-(l_0+l_2)(l_2+l_3)}\\
\notag&\quad\times(\zeta-1)^{(l_0+l_2)(l_1+l_4)-(l_2+l_3)^2-l_3}(\zeta+2)^{-3l_2(l_0+l_2+l_4)-(l_0+l_4)(l_4+1)}\\
\notag&\quad\times(2\zeta+1)^{-l_0^2-l_1(l_0+l_1+3l_2+1)-l_2}u^{\frac 12(l_1-l_3)(l_1+l_3+2l_4-1)+2l_2(l_0+l_2)}\\
&\quad\times 
v^{\frac 12(l_1-l_4)(l_1+l_4+2l_3-1)+2l_2(l_0+l_2)}\tau_0^{l_0+1}\tau_1^{l_1}\tau_2^{l_2}\tau_3^{l_3}\tau_4^{l_4}.
\end{align*}
Recall also the sequence  $Y_k$ defined in \eqref{yre}.

\begin{theorem}\label{trt}
We have
\begin{equation}\label{xtt}\mathbf X\left(\frac{\tau_{l_1l_2l_3l_4}}{\phi_{l_1l_2l_3l_4}}\right)=Y_{k_0}Y_{k_1}Y_{k_2}Y_{k_3}\,t^{(k_0,k_1,k_2,k_3)}, \end{equation}
where 
\begin{equation}\label{klr}
k_0=-l_1-l_2-l_4,\qquad k_1=-l_2,\qquad k_2=-l_1-l_2-l_3,\qquad k_3=-l_2-l_3-l_4.
\end{equation}
\end{theorem}

Note that \eqref{klr} defines a bijection between $\mathbb Z^4$ and the sublattice defined by $\sum k_j\in 2\mathbb Z$.
To prove Theorem \ref{trt} we check that, if one would
define $t^{(\mathbf k)}$ using \eqref{xtt}, all the properties
of Theorem \ref{rt} would hold. This is in principle straight-forward.

Let 
\begin{equation}\label{eql}q_{l_1l_2l_3l_4}=\mathbf X(T_1^{l_1}T_2^{l_2}T_3^{l_3}T_4^{l_4} q)\in\mathbb C(\zeta). \end{equation}
Then $q=q_{l_1l_2l_3l_4}$ solves
 \eqref{py}, with $t$  given by \eqref{tz}
and 
$$(\alpha,\beta,\gamma,\delta)=\left(\frac{l_1^2}2,-\frac{l_4^2}2,\frac{l_3^2}2,\frac{1-l_0^2}2\right).$$
By \eqref{tqf}, this solution can be factored in terms of the
functions $t^{(\mathbf k)}$. To give an example,
\begin{equation}\label{qte} q_{-1,-2,3,-1}=\frac{\zeta(\zeta+2)(\zeta^3+3\zeta^2+3\zeta+5)(5\zeta^3+15\zeta^2+7\zeta+1)}{(2\zeta+1)(5\zeta^3+3\zeta^2+3\zeta+1)(\zeta^3+7\zeta^2+15\zeta+5)}, \end{equation}
where the non-trivial factors can be recognized from
\begin{align*}
t^{(4,1,-1,0)}&=-\frac{4\zeta^{10}(\zeta-1)}{(\zeta+1)(\zeta+2)^{10}(2\zeta+1)}\,(\zeta^3+3\zeta^2+3\zeta+5),\\
t^{(3,2,0,-1)}&=\frac{4(\zeta-1)(\zeta+2)^4(2\zeta+1)}{\zeta^4(\zeta+1)}\,(5\zeta^3+15\zeta^2+7\zeta+1),\\
t^{(4,1,0,-1)}&=\frac{4\zeta^4(\zeta-1)}{(\zeta+1)(\zeta+2)^3(2\zeta+1)^8}\,(5\zeta^3+3\zeta^2+3\zeta+1),\\
t^{(3,2,-1,0)}&=-\frac{4\zeta^{12}(\zeta-1)}{(\zeta+1)(\zeta+2)^{11}(2\zeta+1)^4}\,(\zeta^3+7\zeta^2+15\zeta+5).
\end{align*}

\subsection{Applications}\label{pas}
As an application of Theorem \ref{trt},  we can use  
the results of \S \ref{tls} to study the behaviour of the solutions
\eqref{eql} at the cusps.

\begin{corollary}\label{scc}
Define $\chi(k)$  as $1$ for $k$ odd and $0$ for $k$ even.
Then,
\begin{subequations}
\begin{align}\label{sca}q_{l_1l_2l_3l_4}&=\frac{\zeta^{1+|l_0|\delta_{l_4,0}}(\zeta+2)^{1+\chi(l_1+l_3)}}{(2\zeta+1)^{1+\chi(l_3+l_4)}}\,f(\zeta)\\
\label{scb}&=1+\frac{(\zeta+1)^{1+|l_0|\delta_{l_3,0}}(\zeta-1)^{1+\chi(l_1+l_4)}}{(2\zeta+1)^{1+\chi(l_3+l_4)}}\,g(\zeta),\end{align}
\end{subequations}
with $f$ and $g$ rational functions without zeroes or poles in  $\{0,1,-1,-2,-1/2\}$.
Moreover,
$$
\lim_{\zeta\rightarrow\infty}\zeta^{1+|l_0|\delta_{l_1,0}}{q_{l_1l_2l_3l_4}}
$$
exists and is non-zero.
\end{corollary}

Corollary \ref{scc} encodes the behaviour of the solutions near the singular points of \eqref{py}. For instance, near $t=0$, \eqref{tz} behaves either as $t\sim \zeta$ or as $t\sim(\zeta+2)^3$. The first branch corresponds to $q\sim t^{1+|l_0|\delta_{l_4,0}}$ and the second branch to $q\sim t^{1/3}$ or $q\sim t^{2/3}$, depending on the parity of $l_1+l_3$.

We can also apply Theorem \ref{trt} to deduce new properties of
the functions  $t^{(\mathbf k)}$. For instance, when acting on $\mathcal F_0$, the transformations
 $\{s_0,s_1,s_3,s_4,t_1,t_3\}$ generate the group
$\mathrm S_4\times\mathrm S_2\times\mathrm S_2$.
The property \eqref{xtsx} can be used to obtain a 
 symmetry of the functions $t^{(\mathbf k)}$ under that group,
which agrees with the total set of symmetries discussed in 
\S \ref{sss}. In particular, this proves Proposition~\ref{esp}.

 There are many bilinear relations for tau functions (see e.g.\ \cite{m}) that can be translated to relations for the functions
 $t^{(\mathbf k)}$. To give an example,
\begin{multline*}\frac{\delta^2(\tau_{l_1l_2l_3l_4})\tau_{l_1l_2l_3l_4}}t-\frac{\delta(\tau_{l_1l_2l_3l_4})^2}{t}
-\delta(\tau_{l_1l_2l_3l_4})\tau_{l_1l_2l_3l_4}\\
 \begin{split}&\quad+S(l_0-\alpha_0,l_1-\alpha_1,l_3-\alpha_3,l_4-\alpha_4)\,\tau_{l_1l_2l_3l_4}^2\\
& =(-1)^{l_3+l_4}\ti u\, \tau_{l_1,l_2+1,l_3-1,l_4}\tau_{l_1,l_2-1,l_3+1,l_4}
\end{split}\end{multline*}
holds as an identity in $\mathcal F_0$, where
$$S(l_0,l_1,l_3,l_4)=\frac 1{12}\left(2l_0^2-l_1^2+2l_3^2-l_4^2+6l_0l_3+4l_0+6l_3+2\right).$$
This leads to the following result.

\begin{proposition}\label{nbp}
The functions $t_n=t^{(k_0+n,k_1+n,k_2,k_3)}(\zeta)$, $n\in\mathbb Z$, satisfy the recursion
\begin{multline}\label{nbr}
-(2k_0+2n+1)(2k_1+2n+1)\frac{(\zeta+2)^2}{\zeta^2}\,t_{n+1}t_{n-1}\\
=A
\left(t_n''t_n-\left(t_n'\right)^2\right)+Bt_n't_n+C_n\,t_n^2,
\end{multline}
where 
\begin{align*}
A&=\zeta(\zeta+1)^2(\zeta-1)^2(\zeta+2)(2\zeta+1),\\
B&=2(\zeta+1)^2(\zeta-1)(\zeta^3-3\zeta^2-6\zeta-1),
\end{align*}
and $C_n$ is a polynomial in $k_j$, $n$ and $\zeta$.
\end{proposition}

We refer to \cite{r2} for the explicit expression for $C_n$.
Proposition \ref{nbp} settles some conjectures for polynomials related to
solvable models. The cases $\mathbf k=(0,0,0,0)$ and $(0,0,1,-1)$ prove
\cite[Conj.~1(b)]{bm2}. 
Moreover, the case $\mathbf k=(1,0,0,-1)$ proves
 \cite[Conj.\ 6]{mb}.

Zinn--Justin  \cite{zj} derived recursions for polynomials equivalent to
$t^{(0,2n,0,0)}$ and $t^{(-1,2n+1,0,0)}$. These can also be proved 
 using the
relation to Painlev\'e tau functions. In fact, any 
sequence of the form $t_n=t^{(k_0,k_1+2n,k_2,k_3)}$ satisfies a recursion 
 similar to Proposition~\ref{nbp}.

As another application, it follows from \eqref{qd} that 
$t^{(\mathbf k)}$  satisfies a 
quadratic differential equation. 
This seems to be a novel observation, even in special cases.

\begin{proposition}\label{qdp2}
The function $t=t^{(\mathbf k)}(\zeta)$ satisfies a  differential equation of the form
\begin{equation}\label{qdt}\sum_{i\geq j\geq 0,\ i+j\leq 4}A_{ij}\frac{d^it}{d\zeta^i}\frac{d^jt}{d\zeta^j}=0, \end{equation}
with coefficients $A_{ij}$ that are polynomials in $\zeta$ and $k_0,\dots,k_3$.
\end{proposition}

The coefficients are
 very cumbersome to write down in general. 
To give a concrete example,  consider the case 
$\mathbf k={(0,2n,0,0)}$. It follows from our results  that
$$t^{(0,2n,0,0)}(\zeta)=\left(\frac{\zeta(\zeta+1)}{\zeta+2}\right)^{n(n-1)}f_n\left((2\zeta+1)^2\right), $$
with $f_n$  a polynomial of degree  $\binom n2$,
which is related to the polynomial
$q_n$ of \cite{mb} by
$$q_n(z)=D_n\,z^{n(n+1)}f_{n+1}(z^{-2}), $$
where $D_n$ is a constant, see \eqref{bmq}.
In terms of $f_n(z)$, \eqref{qdt} takes the form
\begin{multline*}
z(z-1)^3(z-9)^3\big(f_n^{(4)}f_n-4f_n^{(3)}f_n'+3(f_n'')^2\big)\\
\shoveleft{+(7z-3)(z-1)^2(z-9)^3\big(f_n^{(3)}f_n-f_n''f_n'\big)}\\
\shoveleft{-2(z-1)(z-9)\big\{(z+1)(z-9)^2n^2+2(z-9)^2n-5z^3+105z^2-483z+351 \big\}f_n''f_n}\\
\shoveleft{+2(z-1)(z-9)\big\{(z+1)(z-9)^2n^2+2(z-9)^2n-z^3+9z^2-111z+135\big\}(f_n')^2}\\
\shoveleft{-\big\{2(z-9)(z^3-39z^2+139z+27)n^2+8(z-9)(3z^2+2z+27)n}\\
\shoveleft{-2z^4+72z^3-876z^2+2184z-1890\big\}f_n'f_n}\\
\shoveleft{-2n(n-1)\big\{(5z-21)(z-9)n^2-(z+15)(z-9)n+z^2+22z+9\big\}f_n^2=0}.
\end{multline*}

As a final remark,
we stress that the functions $t_n^{(k_0,k_1,k_2,k_3)}$ are defined by
 explicit determinants. For instance,  for $n\geq 0$,
\begin{align}\notag t^{(n,n,0,0)}&=\lim_{\substack{x_1,\dots,x_n\rightarrow \xi_0\\y_1,\dots,y_n\rightarrow \xi_1}}
\frac{\prod_{i,j=1}^nG(x_i,y_j)}{\prod_{1\leq i<j\leq n}(y_j-y_i)(x_j-x_i)}\,\det_{1\leq i,j\leq n}\left(\frac{1}{G(x_i,y_{j})}\right),\\
\label{ntd}&=\frac{G(\xi_0,\xi_1)^{n^2}}{\prod_{j=1}^{n}(j-1)!^2}
\,\det_{1\leq i,j\leq n}\left(\frac{\partial^{i+j-2}}{\partial x^{i-1}\partial y^{j-1}}\Bigg|_{x=\xi_0,y=\xi_1}\frac{1}{G(x,y)}\right).
\end{align}
We have proved that
these functions solve the recursion \eqref{nbr}.
This is reminiscent of how the Toda equation
\begin{equation}\label{te}\tau_{n+1}\tau_{n-1}=\tau_n''\tau_n-(\tau_n')^2 
\end{equation}
is solved by Hankel determinants
\begin{equation}\label{tes}\tau_n=\det_{1\leq i,j\leq n}(f^{(i+j-2)}). 
\end{equation}
However, an important difference is that, whereas \eqref{te} is immediately obtained
from \eqref{tes} by applying the Jacobi--Desnanot identity,
that is not the case for \eqref{ntd}. 
It must be combined with 
 the Schr\"odinger equation, which allows us to replace the derivatives in $x$ and $y$ with
 $\zeta$-derivatives (cf.\ \S \ref{cas}).

It should be mentioned that genuine Hankel determinants for tau functions of Painlev\'e VI have been given in \cite{ny2}. These are  
  different
in nature from \eqref{ntd}. It would be interesting to know whether
identities such as \eqref{ntd} are peculiar to our choice of seed solution, or if similar formulas can be found in other situations.

\section{Comparison of notation}
\label{cns}

In this Section, we explain how  $T_n^{(\mathbf k)}$ are related to various polynomials appearing in  \cite{bm1, bm2, bh,fh,h,mb,ras,r,zj}.

\subsection{Polynomials related to the three-colour model}
\label{tcss}

In \cite{r}, we worked  with a symmetric polynomial in $2n+1$ variables, defined by
\begin{multline*}S_n(x_1,\dots,x_n,y_1,\dots,y_n,z)\\
=\frac{\prod_{i,j=1}^nG(x_i,y_j)}{\prod_{1\leq i<j\leq n}(x_j-x_i)(y_j-y_i)}\,
\det_{1\leq i,j\leq n}\left(\frac{F(x_i,y_j,z)}{G(x_i,y_j)}\right),
 \end{multline*}
where $G$ is as in \eqref{g} and
$$F(x,y,z)=(\zeta+2)xyz-\zeta(xy+yz+xz+x+y+z)+\zeta(2\zeta+1).$$
It is not hard to prove that
$$S_n(x_1,\dots,x_{2n+1})=\frac{2^n\prod_{j=1}^{2n+1}(x_j-\zeta)}{1-\zeta}\,T_n^{(0,0,0,-1)}(x_1,\dots,x_{2n+1})
. $$
We can then rewrite 
 the polynomials $P_n$, $p_n$, $y_n$ and $\tilde p_n$
 of \cite[Prop.\ 8.1]{r} in terms of $T_n^{(\mathbf k)}$.
 Namely (recall that $\delta(n)=[n^2/4]$),
\begin{align}
\label{bpr}
P_n(x,\zeta)&=\frac{(-1)^{\left[n/2\right]}\left(\frac\zeta 2+1\right)^{n(n-1)-\delta(n-1)}(x-\zeta)}{(1-\zeta)\zeta^{n(n-1)}(\zeta+1)^{n(n-2)}\left(2\zeta+1\right)^{\delta(n-1)}}\,T_n^{(n,n,0,-1)}(x),\\
\label{pnr} p_n(\zeta)&=\frac{(-1)^{\left[n/2\right]}\left(\frac\zeta 2+1\right)^{n(n-1)-\delta(n-1)}}{(1-\zeta)\zeta^{n(n-1)}(\zeta+1)^{n^2-2n-1}\left(2\zeta+1\right)^{\delta(n)}}\,t^{(n+1,n,0,-1)},\\
\notag y_n(\zeta)&=\frac{(-1)^{\left[n/2\right]}\left(\frac\zeta 2+1\right)^{(n-1)^2-\delta(n-2)}}{2^{[(n+3)/2]}(1-\zeta)\zeta^{n(n-1)}(\zeta+1)^{n^2-2n-1}\left(2\zeta+1\right)^{\delta(n+1)}}\,t^{(n+2,n-1,0,-1)},
\\
\notag\tilde p_n(\zeta)&=
\frac{(-1)^{\left[n/2\right]+1}2^{[(n-1)/2]}\left(\frac\zeta 2+1\right)^{n^2-1-\delta(n)}}{(1-\zeta)\zeta^{n^2-1}(\zeta+1)^{n^2-2n-1}\left(2\zeta+1\right)^{\delta(n-1)}}\,t^{(n,n+1,0,-1)}.
\end{align}
The main result of \cite{r} is that the domain wall partition function for the three-colour model can be expressed in terms of $p_n$ and $\tilde p_n$. The polynomial $P_n$ is related to the domain wall partition function for the 8VSOS model.

\subsection{Polynomials of Bazhanov and Mangazeev}

We will now consider the 
polynomials $\mathcal P_n(x,z)$
 of Bazhanov and Mangazeev \cite{bm1, bm2, mb}, which describe the ground state eigenvalue of Baxter's $Q$-operator for the supersymmetric ($\Delta=-1/2$) periodic XYZ chain of odd length.
In \cite{bm1},  these polynomials are defined up to a factor independent of $x$,
and then normalized  by writing
$$\mathcal P_n(x,z)=\sum_{k=0}^n r_k^{(n)}(z)x^k$$
and requiring that 
$r_n^{(n)}(0)=1$. Since this only determines $\mathcal P_n(x,z)$ 
up to a multiplicative factor $f(z)$ with $f(0)=1$, we make the definition precise by requiring in addition that $\mathcal P_n(x,z)$ is not divisible by any non-constant polynomial in $z$. We can then prove that
$\mathcal P_n$ is related to \eqref{bpr} by
\begin{align*}
\mathcal P_n\left(y,\frac{\zeta}{(\zeta+2)(2\zeta+1)}\right)
&=\left(\frac{2}{(\zeta+2)(2\zeta+1)}\right)^{\delta(n)}\left(\frac{\zeta y+\zeta+2}{\zeta(\zeta+1)}\right)^n
\\
&\quad\times P_n\left(\frac{\zeta(y+2\zeta+1)}{\zeta y+\zeta+2},\zeta\right).
\end{align*}
This is easy to prove up to a $\zeta$-dependent factor, but to
identify that factor we use Theorem \ref{pdet}.

Bazhanov and Mangazeev also introduced the polynomials
\begin{align*}s_n(z)&=r_n^{(n)}(z)=\lim_{x\rightarrow\infty}\frac{\mathcal P_n(x,z)}{x^n}, \\
\bar s_n(z)&=r_n^{(0)}(z)=\mathcal P_n(0,z), 
\end{align*}
which in our notation is
\begin{multline*} \left(\frac{(\zeta+2)(2\zeta+1)}{2}\right)^{\delta(n)}s_n\left(\frac{\zeta}{(\zeta+2)(2\zeta+1)}\right)\\
=\frac{(-1)^{\left[n/2\right]}\left(\frac\zeta 2+1\right)^{n(n-1)-\delta(n-1)}}{\zeta^{n(n-1)}(\zeta+1)^{n(n-1)}\left(2\zeta+1\right)^{\delta(n-1)}}\,t^{(n,n,0,0)},
\end{multline*}
\begin{multline*}
\left(\frac{(\zeta+2)(2\zeta+1)}{2}\right)^{\delta(n)}\bar s_n\left(\frac{\zeta}{(\zeta+2)(2\zeta+1)}\right)\\
=\frac{(-1)^{\left[n/2\right]+1}2^{n-1}\left(\frac\zeta 2+1\right)^{n^2-1-\delta(n-1)}}{\zeta^{n^2-1}(\zeta+1)^{n(n-1)}\left(2\zeta+1\right)^{\delta(n-1)}}\,t^{(n,n,1,-1)}.
\end{multline*}

\subsection{Polynomials of Zinn-Justin}
\label{zss}

In \cite{mb}, Mangazeev and
Bazhanov gave a number of conjectures for eigenvectors
of  the supersymmetric XYZ Hamiltonian on a periodic chain
of odd length.
These involve polynomials $p_n$ (not to be confused with \eqref{pnr}) and $q_n$,
indexed by  $n\in\mathbb Z$, which can conjecturally be used to factorize 
the polynomials $s_n$ and $\bar s_n$. For instance, for $n\geq 0$ it is conjectured that
\begin{equation}\label{spp}s_{2n+1}(y^2)= p_n(y)p_n(-y).\end{equation}

 Zinn--Justin \cite{zj} expressed
 $p_n$ and $q_n$ in terms  of the symmetric
polynomials 
$$H_{2n}(x_1,\dots,x_n,y_1,\dots,y_n)
=\frac{\prod_{i,j=1}^nh(x_i,y_j)}{\prod_{1\leq i<j\leq n}(x_j-x_i)(y_j-y_i)}\,\det_{1\leq i,j\leq n}\left(\frac 1{h(x_i,y_j)}\right), $$
where
$$h(x,y)=1-(3+\zeta_{\text{Z}}^2)xy+(1-\zeta_{\text{Z}}^2)xy(x+y) $$
and $\zeta_{\text{Z}}$ is a parameter (with a subscript to distinguish it from our $\zeta$). It is easy to see that
\begin{equation}\label{thr}T_n(\phi(x_1),\dots,\phi(x_{2n}))\\
=\left(\frac{\zeta(\zeta+1)}{\zeta+2}\right)^{n(n-1)}\,H_{2n}(x_1,\dots,x_{2n}), 
\end{equation}
where
$$\phi(x)=\frac{\zeta}{\zeta+2}\left(1-2(\zeta+1)x\right)$$
and $\zeta_{\text{Z}}=2\zeta+1$.

It follows that, in our notation,
$$p_{n}\left(\frac 1{2\zeta+1}\right)=\frac{(-1)^nC_n(\zeta+2)^{n^2-n-1}}{\zeta^{n^2-2n-1}(\zeta+1)^{n(n-1)}(2\zeta+1)^{n^2+n+1}}\,t^{(-1,2n+1,0,0)}, $$
where $C_n=2^n$ for $n\geq 0$ and $C_n=3^{n+1}/2^{n+2}$ for $n\leq -1$. 
Similarly,
\begin{equation}\label{bmq}q_{n}\left(\frac 1{2\zeta+1}\right)=D_n\left(\frac{\zeta+2}{\zeta(\zeta+1)(2\zeta+1)}\right)^{n(n+1)}t^{(0,2n+2,0,0)}, \end{equation}
where $D_n=1$ for $n\geq -1$ and $D_n=3^{n+2}/2^{2n+3}$ for $n\leq -2$.
The identity \eqref{spp}, and related conjectures from \cite{mb},  can thus be expressed  in terms of   $t^{(\mathbf k)}$. 
We hope to return to these conjectures in the future.

\section{Acknowledgements}
This research has been 
 supported by the Swedish Science Research Council.
  I would like to thank Vladimir Bazhanov, Stefan Kolb, Vladimir Mangazeev, Dmitrii Novikov, Bulat Suleimanov and Paul Zinn-Justin for interesting discussions and correspondence. I am also grateful to Nikolai Reshetikhin who  suggested that I write the present summary of the work described in  \cite{r1a,r1b,r2}.

 \end{document}